\documentclass[11pt]{article}
\usepackage[utf8]{inputenc}
\usepackage{amssymb,amsmath}
\usepackage{bm} 
\usepackage{booktabs} 
\usepackage{array}
\usepackage{float}

\usepackage{latexsym}
\usepackage{graphicx}
\usepackage{color}
\usepackage{datetime}
\usepackage[nosort]{cite}
\usepackage{verbatim}
\usepackage{enumerate}
\usepackage{chngpage} 
\usepackage{mathrsfs}
\usepackage{euscript}
\usepackage{psfrag}
\usepackage{subcaption}
\usepackage{tikz}

\usepackage{datetime}

\usepackage[nosort]{cite}
\usepackage{chngpage} 
\usepackage{setspace}
\usepackage{tensor}
\usepackage{physics}

\def\be{\begin{equation}}
\def\ee{\end{equation}}
\def\bea{\begin{eqnarray}}
\def\eea{\end{eqnarray}}

\usepackage{mciteplus}

\usepackage[colorlinks=true,      linkcolor=blue,      urlcolor=blue,      
            filecolor=blue,      citecolor=blue,       pdfstartview=FitH,     
						pdfpagemode=UseNone,      bookmarksopen=true]{hyperref}  
\usepackage[all]{hypcap}     

\definecolor{cardinal}{rgb}{0.6,0,0}
\definecolor{darkgreen}{rgb}{0,0.4,0}
\definecolor{golden}{rgb}{0.92, 0.7, 0}
\definecolor{midnight}{rgb}{0, 0, 0.5}
\definecolor{darkblue}{rgb}{0, 0, 0.7}
\definecolor{purple}{rgb}{0.5, 0, 0.5}

\def\oneone{\rlap 1\mkern4mu{\rm l}}

\def\Neql#1{{\cal N}\!=\!{#1}}
\def\coeff#1#2{\relax{\textstyle \frac{#1}{#2}}\displaystyle}

\def\IR{\mathbb{R}}

\def\IT{\mathbb{T}}

\def\ZZ{\mathbb{Z}}

\def\cB{{\cal B}}

\def\cF{{\cal F}}

\def\cO{{\cal O}}
\def\cP{{\cal P}}

\def\nBPS#1{$\frac{1}{#1}$-BPS}

\begin{document}
 \numberwithin{equation}{section} 

\phantom{AAA}
\vspace{-10mm}

\begin{flushright}
%
%
\end{flushright}

\vspace{1.9cm}

\begin{center}

{\huge {\bf Microstate Geometries}}\\
\vspace{0.4cm}

\vspace{1.0 cm}

{\large{\bf { Iosif Bena$^{1}$  and  Nicholas P. Warner$^{1,2,3}$}}}

\vspace{0.5cm}

$^1$Institut de Physique Th\'eorique, \\
Universit\'e Paris Saclay, CEA, CNRS,\\
Orme des Merisiers, Gif sur Yvette, 91191 CEDEX, France \\[12pt]

\centerline{$^2$Department of Physics and Astronomy}
\centerline{and $^3$Department of Mathematics,}
\centerline{University of Southern California,} 
\centerline{Los Angeles, CA 90089, USA}

 \vspace{5mm} 
{\footnotesize\upshape\ttfamily iosif.bena @ ipht.fr,   warner @ usc.edu} \\

\vspace{1.5cm}
 
\textsc{Abstract}
\end{center}

\begin{adjustwidth}{3mm}{3mm} 
 
We review the 20-year history of the {\it Microstate Geometry Programme} and the essential role that supergravity has played, and will continue to play, in the description of black-hole microstructure.
\vspace{-1.2mm}
\noindent

\end{adjustwidth}

\vfill

{{\it Contribution to the volume ``Half a Century of Supergravity,'' eds.~A.~Ceresole
and G.~Dall'Agata.}}

\thispagestyle{empty}
\newpage


\tableofcontents


\section{Introduction}
\label{sec:Intro}

 Our current understanding of black holes strongly suggests that supergravity, in many of its forms, has a very important role to play in our fundamental understanding of the universe.  Long ago, the singularity theorems told us  that general relativity predicts its own failure because  singularities are inevitable consequences of the theory \cite{Penrose:1964wq}.   The general belief was that one would need some quantum theory of gravity to resolve these singularities, and for many years it was believed that such a resolution was always going to be confined to the Planck scale.   However, we now understand that the event horizon of a black hole is almost as pathological as the singularity, and that the quantum resolution of a black hole must happen at the horizon scale.
  
 The Hawking information paradox  \cite{Hawking:1976ra,Hawking:1982dj} was greatly sharpened by the small corrections theorem  \cite{Mathur:2009hf}  and this tells us that if one has a smooth, geometric horizon then the Hawking process will necessarily lead to either a huge violation of quantum unitarity, or extreme levels of non-locality in quantum field theory.  Resolving this dilemma has been a major focus in theoretical physics for decades and it has led a significant number of researchers to believe that the event horizon of general relativity theory must  simply be an artifact of the failure of an effective field theory.  More specifically,  there has to be a more fundamental quantum theory of black holes that can describe their microstucture and recast Hawking evaporation as a unitary process.  Such a description of a black hole must necessarily be horizonless so that distant observers can access all the microstates,  enabling measurements of interior structure and, through this, the preservation of unitarity. This idea is the core principle of the {\it fuzzball proposal}.
  
General Relativity (GR) would then emerge as an effective classical limit of the more fundamental theory, and  horizons  reflect  the inability of GR to describe the microscopic details of  black holes:  GR merely provides a thermodynamic, or hydrodynamic, ensemble-average description of those details.   The startling thing about this new perspective is that it means black holes are necessarily quantum objects {\it at the horizon scale}. There are a several arguments (see, for example, \cite{Mathur:2008kg}) as to why this is must occur, and we will not review them here\footnote{For reviews, see \cite{Bena:2022ldq,Bena:2022rna}}, but suffice it to say, the horizon-scale quantum properties emerge as a collective effect of the truly vast number of microstates  that make up a black hole.   

Since string theory is the only really viable theory of quantum gravity, it plays the dominant role in the analysis of quantum black holes and underlies the structure of fuzzballs. However, the string scale is vastly smaller than the horizon scale of macroscopic black holes, and so the first step in describing the large-scale, collective effects of a fuzzball must involve the massless sector of string theory, namely the supergravity limit. Put differently, a fuzzball must have coherent states that are well-described by the geometry and fluxes of supergravity solutions.  Finding the smooth, horizonless supergravity solutions that describe the coherent microstructure of black holes defines the {\it microstate geometry programme},  and this chapter reviews the state of this particular craft.    A microstate geometry is a supergravity soliton:  a smooth, horizonless supergravity solution that ``looks like'' a black hole from a distance.  Ideally, one would like such a solution to conform closely to that of a black hole except that the microstate geometry ``caps off'' smoothly, at high red-shift, just above the horizon scale   

The huge entropy of a black hole  not only implies that it must have an insane number of microstates, but also  strongly suggests that there must be a proportionally insane number of coherent expressions of those states.  As we will discuss, depending on dimension and duality frame, microstate geometries can be captured  and  studied using a range of supergravity theories, and many of the solutions can be mapped to the holographically dual field theories that underlie the microstructure of black holes.  At its most basic level, one may think of supergravity as another low-energy, effective description of black holes but supergravity is  hugely more effective than general relativity theory because it has many more degrees of freedom and can realize a vast array of smooth, solitonic, horizonless solutions that are demonstrably dual to black-hole states. The microstate geometry programme has thus given extensive computational support for the fuzzball paradigm, and has helped elevate fuzzballs to the most  substantial and well-formulated resolution to the information problem.

Because of the vast range of ``no-go'' theorems for gravitational solitons in $3+1$ dimensions, our story starts in five dimensions.

\section{The Genesis of microstate geometries: Supergravity in five dimensions}
\label{sec:5d}

The first macroscopic black hole whose entropy was described in string theory is the supersymmetric, three-charge black hole in five dimensions \cite{Strominger:1996sh}. The full ten- or eleven-dimensional solution corresponding to this black hole can be reduced to a solution of $U(1)^3$, five-dimensional, ungauged supergravity with three electric charges. When the three electric charges are equal, this theory reduces to minimal ungauged supergravity.

Amazingly, these rather simple supergravity theories  allow one to build horizonless microstate geometries corresponding to these black holes. The metric of the black-hole solution solution can be written as
\begin{equation}
ds^2 = - Z^{-2} (dt+k)^2 + Z(dr^2 + r^2 d \Omega_3^2)~,~~~{\rm with}~~~Z=1+{Q/r^2}\,.
\label{BH-general}
\end{equation}
The warp factor, $Z$,  has a pole at $r=0$, which corresponds to the location of the event horizon.   To create a microstate geometry, one must build a smooth, horizonless  solution with exactly the same asymptotics at infinity.  

To see how one might achieve this this, one should remember that, in supergravity, the sources for the potential and warp factor, $Z$, can have two very different contributions:
\begin{equation}
\nabla^2 Z ~=~  Q_{\rm localized} \delta(r)~+~  *_4 \theta \wedge \theta \,. 
\label{genericZth}
\end{equation}
The $\delta$-functions are point charges and inevitably lead to singularities and horizons.  The other possibility is the sourcing of electric charges through magnetic fluxes and Chern-Simons terms.   In (\ref{genericZth}), supersymmetry requires $\theta$ to be a magnetic, self-dual, harmonic two-form  on the four-dimensional spatial ``base space'' \cite{Gauntlett:2002nw}. The operator $*_4$ is the Hodge duality on this base.     Most significantly, harmonic fluxes can be sourced smoothly from cohomological fluxes threading non-trivial $2$-cycles.

In general, supersymmetry also requires the metric on the base manifold to be hyper-K\"ahler.   For the black-hole solution, \eqref{BH-general}, the base is $\IR^4$, there is no non-trivial topology and $\theta \equiv 0$. To avoid a horizon one must replace the $\delta$-functions, with cohomological  magnetic flux.  Fortunately, such non-trivial cohomology is typically a structural feature of hyper-K\"ahler manifolds.  Since we want the same asymptotics at infinity as the black hole, we are led to requiring a  hyper-K\"ahler base that is asymptotic to $\IR^4$. 

At this point the programme appears to come to a very abrupt end:  $\IR^4$ is  the {\it only} regular,  hyper-K\"ahler space that is asymptotic to $\IR^4$.  It seems there is  no topology and so no smooth cohomological flux.  However, supergravity is smarter. One can actually build completely regular, five-dimensional, {\it Lorentzian} solutions starting from ``ambi-polar'' hyper-K\"ahler bases that, by definition, have regions of signature ($+,+,+,+$) and regions of signature ($-,-,-,-$).   These four-dimensional geometries are singular, but the five-dimensional geometry is not only Lorentzian and smooth but can also have highly non-trivial topology.

The simplest  examples of this come from extending the idea of Gibbons-Hawking (GH) spaces.  One takes a base metric of the form:
\begin{equation}
ds_{GH}^2 = V^{-1}(d \psi+ A)^2 + V (d \vec x\cdot d\vec x )\,, \qquad {\rm with} \qquad \vec \nabla \times \vec A = \vec \nabla V \,,
\label{GHmet}
\end{equation}
where $\vec x \in \IR^3$ and $A = \vec A \cdot d \vec x$.   This metric is hyper-K\"ahler and  it is the flat metric on $\IR^4$ if $V =\frac{1}{r}$, where $r \equiv \sqrt{\vec x\cdot \vec x}$.  The metric is asymptotic to $\IR^4$ if  $V \sim \frac{1}{r}$ as $r \to \infty$.    

The standard GH metric has 
\begin{equation}
V~=~  \sum_{i= 1}^ N \, \frac{q_i}{| \vec x - \vec a_i |} \,, 
\end{equation}
for some charges, $q_i \in \ZZ$.  To be a Riemannian (positive-definite) metric one must impose $q_i \ge 0$, however one can obtain ambi-polar hyper-K\"ahler metrics by dropping this condition and simply taking $q_i \in \ZZ$.  Such a metric is the asymptotic to $\IR^4$ if $\sum_i q_i =1$.   Allowing $q_i \in \ZZ$, means that   $V$ can vanish and the metric (\ref{GHmet}) is extremely singular on the hypersurfaces defined by $V=0$. 

The $\psi$-fiber in (\ref{GHmet})  pinches off at $\vec x = \vec a_i$, and the homology $2$-cycles, $\Delta_{ij}$, can be exhibited by fibering the $\psi$-circle over any curve between any two points, $\vec a_i$ and $\vec a_j$.  There is, however, the danger that the singular surfaces, $V=0$,  could decompactify some of these putative homology cycles.

The miracle of five-dimensional supergravity is that, despite the singularities of the base metric, the  five-dimensional geometry can be rendered smooth, and the homology cycles are indeed defined by the $\Delta_{ij}$.  The five-dimensional metric is:
\begin{equation}
ds^2 = - Z^{-2} (dt+k)^2 + Z ds_{GH}^2  =  - Z^{-2} (dt+k)^2 + Z V^{-1}(d \psi+ A)^2 + Z V (d \vec x\cdot d\vec x )\,.
\label{GH-full}
\end{equation}
For this to be smooth, three miracles need to occur. First, at the interface between ($+,+,+,+$) and ($-,-,-,-$) regions, where $V$ changes sign, the warp factor, $Z$, must also change sign  {\it by going through a pole},  in such as way that the coefficient, $ZV$, of the metric along $\IR^3$ remains smooth. This, however, creates new problems: The coefficient of the $(d\psi+A)^2$ term blows up like $V^{-2}$.  The only way that this divergence, and the sub-leading  $V^{-1}$ divergence,  can be canceled is via the $- k^2$ term coming from $-(dt+k)^2$ term in (\ref{GH-full}).    It was shown in  \cite{Bena:2005va,Berglund:2005vb,Bena:2007kg}  that this indeed happens, along with similar ``miraculous'' cancellations in the electromagnetic fields. Thus,  five-dimensional supergravity does indeed contain a very large family of multi-bubble solutions constructed using ambi-polar Gibbons-Hawking base spaces. Note that these miraculous cancelations are not a consequence of supersymmetry. One can construct non-supersymmetric solutions by replacing the hyper-K\"ahler four-dimensional base manifold with any Ricci-flat space \cite{Bena:2009fi}, and one can obtain smooth solutions of five-dimensional supergravity even when this Ricci-flat space changes signature \cite{Bena:2009qv,Bobev:2009kn}.

Another very practical miracle that occurs in the construction of five-dimensional microstate geometries is that, once one has chosen the hyper-K\"ahler metric, the remaining BPS equations, governing the electromagnetic fields, and the other pieces of the metric, are {\it linear} \cite{Bena:2004de}!   The equations actually reduce to those of four-dimensional electromagnetism on the base.  This makes the construction of five-dimensional microstate geometries largely algorithmic.   It is important to stress that because the equations of motion involve Einstein's equations for the metric, and Chern-Simons interactions for the Maxwell fields, the BPS equations could have been hopelessly non-linear.  They are not, and, as we will discuss, this seems to be a feature of our approach to microstate geometries, and has led to a mathematical part of the {\it Themelion conjecture.}

Subsequent work \cite{Bena:2006kb, Bena:2007qc} showed that one could create smooth, horizonless geometries that have ``scaling behavior'' and ``cap off''  at arbitrary high redshift.There is now an extensive range of such bubbling solutions, both for supersymmetric black holes. 
\cite{
Bena:2005va,
Berglund:2005vb,
Bena:2007kg,
Giusto:2004kj,
Bena:2006is,
Bena:2008wt,
Bena:2010gg,
Bena:2011dd, 
Bena:2013dka, 
Bianchi:2016bgx, 
Bianchi:2017bxl, 
Heidmann:2017cxt, 
Bena:2017fvm, 
Avila:2017pwi,
Tyukov:2018ypq,
Warner:2019jll,
Rawash:2022sum} 
and also for extremal non-supersymmetric black holes
\cite{Goldstein:2008fq,
Bena:2009ev,
Bena:2009en,
DallAgata:2010srl,
Vasilakis:2011ki,
Heidmann:2018mtx} \footnote{Many of these solutions preserve a tri-holomorphic $U(1)$ isometry and can be formally compactified to solutions of four dimensional ungauged supergravity, that belong to the class of multi-center solutions of \cite{Denef:2000nb,Denef:2002ru, Bates:2003vx,Balasubramanian:2006gi}.}.

Semi-classical quantization of these solutions then revealed that there is a maximal depth \cite{deBoer:2008zn, deBoer:2009un,Martinec:2015pfa,Li:2021gbg}, or red-shift, for the throat and that excitations within the cap have  an energy gap that perfectly matches with the black-hole CFT.

The fact that one can make smooth solitonic solutions to supergravity was already a remarkable triumph: an important proof-of-principle that string theory and its low-energy limit, supergravity, could avoid a plethora of ``no-go'' theorems and describe at least some of the horizon-scale structure that is essential to resolving the information paradox.   Central to this is the discovery of a ``support mechanism'' for horizon-scale structure:  Normal matter immediately collapses into a black hole, and this lies at the core of  black hole uniqueness, as well as the soliton  ``no-go,''  theorems.  The non-trivial topology and cohomological fluxes inherent in microstate geometries were subsequently shown \cite{Gibbons:2013tqa} to be the {\em only} possible semi-classical support mechanism, both in five {and in six} dimensions \cite{deLange:2015gca}.  This also means that any discussion of horizon-scale microstructure that includes gravitational back-reaction must necessarily involve microstate geometries.  Note that the argument of \cite{Gibbons:2013tqa} does not rely on supersymmetry (although at the time only supersymmetric solutions were known): We now have extensive examples of multi-bubbled microstate geometries that have the same charges as non-extremal non-BPS black holes 
\cite{Bah:2021owp,
Bah:2021rki,
Heidmann:2021cms,
Bah:2022yji,
Bah:2022pdn,
Heidmann:2022zyd,
Bah:2023ows}.

The next question was how much of the microstructure are these solutions able to capture.  The holy grail is the entropy of the BPS black hole, $S \sim Q^{3/2}$.  Simple counting of the quantized moduli spaces and partitioning of fluxes over bubbles (homology cycles) \cite{Bena:2006is,deBoer:2008zn, deBoer:2009un} suggested that five-dimensional microstate geometries could account for only about $S \sim Q^{1/2}$.    At a technical level, the five-dimensional microstate geometries simply have too many isometries to see detailed microstructure, and with hindsight we now know that, roughly speaking, bubbling geometries are primarily capturing the super-selection sectors of the theory that is dual to the black hole, but not the most generic states that live in these sectors.

To see more of the microstructure, one must reduce the amount of symmetry.   An important guide to doing this came from the earlier work on the two-charge problem and the fact that {\it supertubes} \cite{Mateos:2001qs,Emparan:2001ux,Lunin:2001fv,Lunin:2002iz,Palmer:2004gu, Rychkov:2005ji} could capture a lot of entropy: 
\begin{equation}
S_{\rm supertubes} ~=~ 2 \pi \sqrt{N_1 N_5} ~\sim~ Q
\end{equation}
where $N_1$ and $N_5$ are the number of underlying branes.  Indeed this entropy comes from quantizing solutions parameterized by several arbitrary continuous functions of one variable that describe the shape and charge densities of the supertube.

There are two fundamental ways to leverage the enormous entropy of supertubes into constructing large numbers of three-charge microstate geometries. The first, which we will not review here, is to place supertubes parameterized by arbitrary functions inside bubbling solutions \cite{Bena:2008nh,Bena:2008dw,Bena:2010gg}. The second is to construct {\it superstrata} by adding momentum waves to the supertube. These waves break one of the isometries that are used to reduce ten-dimensional supergravity to five dimensions, and hence are not described by smooth solutions of five-dimensional supergravity. To describe these waves in terms of smooth horizonless geometries one needs to use six-dimensional supergravity.

\section{Superstrata and holography: Supergravity in six dimensions}
\label{sec:6d}

The graviton multiplet  of six-dimensional minimal supergravity involves both a graviton and a self-dual tensor gauge field, $B_{\mu\nu}^+$.    To describe generic three-charge objects one must add to this an (anti-self-dual) tensor multiplet, which also contains a scalar ``dilaton,'' $\phi$.  If one reduces this to five dimensions, each tensor fields give rise to a five-dimensional vector and the third  vector field comes, via the Kaluza-Klein mechanism, from the graviton.  Thus all of the earlier solutions of five-dimensional supergravity can easily be uplifted to six dimensions, but now there is the Kaluza-Klein circle, $y$, on which there can be dynamics. 

The most general solution with four supersymmetries was analyzed in \cite{Gutowski:2003rg,Cariglia:2004kk}, and in particular, the metric must take the form
\begin{equation}
d s^2_{6} ~=~ -\frac{2}{\sqrt{\cP}}\,(d v+\beta)\,\Big[d u+\omega + \frac{\mathcal{F}}{2}(d v+\beta)\Big]+\sqrt{\cP}\,d s^2_4\,,
\end{equation}
where
\begin{equation}
u~=~ \coeff{1}{\sqrt{2}}\, (t-y) \,,\qquad v ~=~ \coeff{1}{\sqrt{2}}\, (t+y)  \,, 
\end{equation}
are null coordinates and $ y  \equiv~ y+ 2 \pi R_y $. Supersymmetry requires that everything is independent of $u$, and we will take the metric, $d s^2_4$, on the spatial base, $\cB$, to be a $v$-independent and hyper-K\"ahler  metric\footnote{There is a more general option described in   \cite{Gutowski:2003rg,Cariglia:2004kk}, in which $d s^2_4$ can depend on $v$ and be ``almost hyper-K\"ahler.'' As a practical matter, there are very few known solutions of this form and so we have simplified the form of the metric.}.  The Kaluza-Klein vector, $\beta$, must then also be independent of $(u,v)$ and is required to satisfy:
 \begin{equation}
\Theta_3 ~\equiv~ d \beta = *_4 d\beta\,,
\label{eqbeta}
 \end{equation}
where $*_4$ is the four-dimensional Hodge dual in the base metric.   Indeed, the six-dimensional tensor gauge fields can be decomposed into potentials, $Z_I$, and $2$-form fluxes, $\Theta^{(I)}$, whose components are only along the base space.  However, these fields are all allowed to depend on $v$.  

The BPS equations on these fields can then be written as \cite{Bena:2011dd, Giusto:2013rxa}:
 \begin{equation}
  \begin{aligned}
 *_4 \mathcal{D} \dot{Z}_1 =  \mathcal{D} \Theta_2\,,\quad \mathcal{D}*_4\mathcal{D}Z_1 = -\Theta_2\wedge d\beta\,,\quad \Theta_2=*_4 \Theta_2\,, \\ 
 *_4 \mathcal{D} \dot{Z}_2 =  \mathcal{D} \Theta_1\,,\quad \mathcal{D}*_4\mathcal{D}Z_2 = -\Theta_1\wedge d\beta\,,\quad \Theta_1=*_4 \Theta_1\,,  \\
 *_4 \mathcal{D} \dot{Z}_4 =  \mathcal{D}  \Theta_4\,,\quad \mathcal{D}*_4\mathcal{D}Z_4 = -\Theta_4\wedge d\beta\,,\quad \Theta_4=*_4 \Theta_4\,.
   \end{aligned}
 \label{BPSlayer1}
\end{equation}
where the dot denotes $ \frac{\partial}{\partial v}$,  $\mathcal{D}$ is defined by
\begin{equation}
\mathcal{D} \equiv \tilde d - \beta\wedge \frac{\partial}{\partial v}\,,
\end{equation}
and $\tilde d$ denotes the exterior differential on the spatial base, $\cB$.  How the $(Z_I,  \Theta^{(I)})$ can be  assembled into the six-dimensional tensor gauge fields can be found in \cite{Bena:2011dd, Giusto:2013rxa,Bena:2017geu}.

One should also note we have added another tensor multiplet ($Z_4,  \Theta_4$) and these multiplets, for historical reasons\footnote{This comes from the five-dimensional formulation in which $I=3$ is a gauge field that has been uplifted to the Kaluza-Klein vector, $\beta$, as indicated in (\ref{eqbeta}).}, are indexed by $I=1,2,4$.  One can add further tensor multiplets, but superstrata require  three tensor fields:  One is sourced by D1 branes, another by the D5 branes, and the new one, ($Z_4,  \Theta_4$), will encode NS sector fluxes { sourced by equal amounts of F1 strings and NS5 branes}.  This extra multiplet is essential to complete the stringy solution \cite{Giusto:2011fy}: In addition to the D1 and D5 charges and densities, we need a third type of momentum charge and density to be carried by certain open strings  stretched  between D1  and  D5 branes.  The collective effects of these strings source  ($Z_4,  \Theta_4$).

As one should expect for a BPS solution, the metric warp factor and scalars are expressed in terms of the warp factors:
\begin{equation}
\cP   ~=~     Z_1 \, Z_2  -  Z_4^2 \,, \qquad  e^{2\Phi}~=~ \frac{Z_1^2}{\cP}\,,  \qquad     C_0~=~\frac{Z_4}{Z_1}   \,.
\label{Psimp}
\end{equation}
Finally, the remaining metric quantities, $\omega$ and $\mathcal{F}$, are determined using:
\begin{equation}
\begin{aligned}
\mathcal{D} \omega + *_4  \mathcal{D}\omega & ~=~ Z_1 \Theta_1+ Z_2 \Theta_2  - \mathcal{F}\Theta_3 -2\,Z_4 \Theta_4 \,, \\
 *_4\mathcal{D} *_4\!\big(\dot{\omega} - \coeff{1}{2}\,\mathcal{D} \mathcal{F} \big)&~=~\dot{Z}_1\dot{Z}_2+Z_1 \ddot{Z}_2 + Z_2 \ddot{Z}_1 -(\dot{Z}_4)^2 -2 Z_4 \ddot{Z}_4-\coeff{1}{2} *_4\!\big(\Theta_1\wedge \Theta_2 - \Theta_4 \wedge \Theta_4\big) \\
 &~=~\partial_v^2 (Z_1 Z_2 - {Z}_4^2)  -(\dot{Z}_1\dot{Z}_2  -(\dot{Z}_4)^2 )-\coeff{1}{2} *_4\!\big(\Theta_1\wedge \Theta_2 - \Theta_4 \wedge \Theta_4\big)\,.
\end{aligned}
\label{BPSlayer2}
\end{equation}

The first thing to note is that, once we have chosen the base metric (which we now know can be ambi-polar), all the equations for  $Z_I,  \Theta_I, \omega$ and $\cF$ are linear, and in principle the construction of solutions is algorithmic.  In practice it can be challenging to solve the Poisson equations and find smooth solutions.

The seed for {\it superstratum construction} is the supertube solution \cite{Emparan:2001ux,Lunin:2001fv,Lunin:2002iz}.  This has an extremely simple metric provided one works in spheroidal coordinates.  Specifically, one can write the flat metric on $\IR^4$ as:
\begin{equation}
ds_4^2 ~=~ \Sigma\, \bigg( \frac{dr^2}{(r^2 + a^2)} ~+~ d \theta^2 \bigg)  ~+~ (r^2 + a^2) \sin^2 \theta \, d\varphi_1^2 ~+~ r^2  \cos^2 \theta \, d\varphi_2^2  \,, 
\label{bipolmet}
\end{equation} 
where $\Sigma  \equiv (r^2 + a^2 \cos^2 \theta)$.  The $k$-times wound supertube is then given by:
\begin{equation} 
\beta ~=~\frac{2\,k R a^2 }{\Sigma} \, ( \sin^2 \theta \, d\varphi_1-   \cos^2 \theta \, d\varphi_2 ) \,,  \qquad \omega~=~\frac{2\,k R a^2 }{\Sigma} \, ( \sin^2 \theta \, d\varphi_1 +   \cos^2 \theta \, d\varphi_2 )  \,.
\label{betaform3}
\end{equation} 
and
\begin{align}
Z_1 &~=~ 1 ~+~ \frac{Q_1}{\Sigma} \,, \qquad Z_2 ~=~ 1 ~+~ \frac{Q_2}{\Sigma} \,, \qquad \mathcal{F} ~=~ 0   \,, \qquad Z_4 ~=~ 0 \,,\nonumber  \\
\Theta_3 &~=~ d\beta \,,   \qquad  \Theta_I ~=~ 0\,, \ \ {I =1,2,4}   \,,
\label{STsol1}
\end{align} 
The constants $Q_1$, $Q_2$  are harmonic sources that encode the D1 and the D5 charges of the supertube.

Superstrata are obtained by superposing oscillating waves of fluxes on the supertube.   That is, one solves (\ref{BPSlayer1}) for oscillating, $v$-dependent fluxes and potentials, and feeds those solutions into   (\ref{BPSlayer2}) and solves for the rest of the metric.  One must then, if possible, find homogeneous solutions that make the  fluxes and geometry completely smooth and horizonless.   The linearity of the BPS equations, and the ability to superpose solutions,  is utterly critical to this process and to the enumeration of superstrata.  

Achieving this and the construction of the most general family of superstrata was a long and complex process,  completed through the effort of several groups  (See, for example, \cite{
Giusto:2004kj,
deLange:2015gca,
Bena:2015bea,
Bena:2016agb,
Bena:2016ypk,
Bena:2017xbt,
Bakhshaei:2018vux,
Ceplak:2018pws,
Heidmann:2019zws,
Mayerson:2020tcl,
Ganchev:2021iwy}.)  
There were several important aspects of this process.
First, one finds that one must ``coiffure'' the Fourier modes.  That is, to obtain  smooth solutions, one must impose some algebraic constraints that relate some of the Fourier coefficients to one another.  This was subsequently understood through the holomorphic structure of superstrata \cite{Heidmann:2019xrd}.   Even more importantly, many aspects of the construction, including the coiffuring relations, were tested through precision holography of the D1-D5 system.  Indeed,  the development of superstrata propelled the holography of the D1-D5 CFT duality into one of the thoroughly and extensively understood holographic field theories that we now know. (See, for example, \cite{
Giusto:2015dfa,
Galliani:2016cai,
Bombini:2017sge,
Bombini:2019vnc,
Tian:2019ash,
Tormo:2019yus,
Giusto:2019qig,
Bena:2019azk,
Giusto:2019pxc,
Hulik:2019pwr,
Giusto:2020mup,
Ceplak:2021wzz,
Rawash:2021pik}.)

Finally, it is also possible to construct strongly non-BPS versions of superstrata, known as {\it microstrata} \cite{Ganchev:2021pgs,Ganchev:2021ewa,Ganchev:2023sth,Houppe:2024hyj}, and the breakthrough here was made using yet another form of supergravity:  $\Neql{4}$ gauged supergravity in three dimensions.

It was also becoming clear that the basic class of superstrata, using tensor multiplets, was missing some of the excitations needed to resolve some of the singularities in six-dimensional supergravity \cite{Bena:2022sge}. Indeed, perturbative world-sheet methods had greatly advanced, and could be used to enumerate the perturbative modes  corresponding to the most general possible superstratum excitations \cite{Martinec:2022okx,Ceplak:2022wri}.   This led to the extension of basic superstrata to include vector multiplets and the construction of ``vector superstrata,'' \cite{Ceplak:2022wri,Ceplak:2022pep,Ceplak:2024dbj}.  Remarkably, the BPS equations of this even more general class of superstrata were, once again, shown to have a linear structure.  More importantly, this work showed that the superstratum story was now complete:  superstrata now exactly match with a precisely-known family  of CFT excitations called the ``supergraviton gas,'' which is generated  by the lowest, non-trivial modes in the untwisted sector of the CFT \cite{Shigemori:2019orj,Mayerson:2020acj}.  

 We thus had all the black-hole microstructure that could be described by superstrata, but when these states were counted \cite{Shigemori:2019orj,Mayerson:2020acj},  they yielded  an entropy of, at most, $S \sim (N_1 N_5)^{1/2} N_p^{1/4} \sim Q^{5/4}$, which is parametrically short if the black-hole entropy, $ 2 \pi (N_1 N_5)^{1/2} N_p^{1/2}\sim Q^{3/2}$.    

We knew what the superstrata were missing:  the twisted sectors of the underlying CFTs.  These sectors contain the supersymmetric lowest-mass-gap states of the black-hole CFT, but despite some valiant efforts \cite{Bena:2016agb,Shigemori:2022gxf}, only very restricted sectors are visible in six-dimensions.

{
\subsection{Fortuity, monotony and microstate geometries}
\label{ss:FMMG}

An extremely interesting body of ideas has been emerging from the study of BPS states purely within CFT's (see, for example, \cite{Chang:2024zqi,Chang:2025rqy}).   The idea is to classify states according to how they behave as a control parameter, like the number of branes, $N$, increases, or, in the D1-D5 system, as one dials up or down the values of $N_1$ and $N_5$.  {\it Monotonous BPS states} exist at all values of $N$, while {\it fortuitous BPS states} only exist  for a limited set of values of $N$.  Typically, the existence of the fortuitous BPS states relies on some special cancelations that happen only at certain values of $N$, and that disappear as $N$ is increased.  At larger values of $N$, a fortuitous   BPS state might cease to be exist, but then new fortuitous states, that did not exist at lower $N$, appear.

The suggestion is that monotonous states should be more naturally related, through holography, to BPS supergravity solutions because $N$ typically appears as a supergravity charge that can be dialed, at will, without changing other aspects of the solution. 

Certainly superstrata, which are holographically dual to the supergraviton gas states, fit the pattern of monotone states.  As one increases  $N_1$ and $N_5$ by units of one, the size of the underlying supertube increases monotonically, and the profile of the momentum-carrying wave remains similar. In the CFT, superstrata are obtained by acting with powers of operators like $L_{-1}-J^3_{-1}$, on certain CFT strands. If the strands become longer or shorter the states dual to the superstrata continue to exist. 

On the other hand, the D1-D5 states that count the black-hole entropy come from 1-5 strings that stretch between co-prime numbers of D1 and D5 branes. The momentum of these strings is  quantized in units of $1/(N_1 N_5 R_y)$. Adding momentum carried by these strings corresponds, in the CFT, to acting with highly twisted operators like $L_{-\frac{1}{N_1 N_5}}$.  It is clear that if one changes $N_1$ and/or $N_5$ by just one unit, to make them both even, the momentum modes will be quantized in units of at least $4/(N_1 N_5 R_y)$, and most of the states will disappear (and the entropy will drop by a factor of two).   Hence, these states are fortuitous. 

Na\"ively, it may seem that supergravity geometries and fortuity are somewhat inconsistent paradigms.  However, this is simply not true, and, in particular,   {\it multi-centered} five-dimensional microstate geometries exhibit gravitational analogues  of fortuity.  The important point is that the charges are not the control parameters of the multi-centered geometries: one cannot simply add a unit of one species of charge.  In the multi-centered solutions, the control parameters are the quantized magnetic fluxes, and increasing a single flux can have multiple consequences: 
\begin{itemize} 
\item[(i)]  It generically changes all the charges in a complicated manner because the charges are the result of the Chern-Simons interactions between all the fluxes, \item[(ii)]  It can radically change the geometry of the solution,  particularly the depth of any black-hole-like throat, thus creating a radical change in the energy gap. 
This comes about because the depth of the throat is an extremely delicate and sensitive function of the moduli and the magnetic fluxes because of a set of constraints know as the bubble equations \cite{Bena:2005va,Bena:2007kg}.   
\item[(iii)]   The solution can cease to exist:  no matter how one varies the moduli, it is possible that there will be no physical solutions to the bubble equations for some choices of fluxes. 
 \item[(iv)] Changing some of the moduli can deform the theory and shift the definitions of the charges so that a particular multi-center BPS solution may cease to exist (see, for example, \cite{Bossard:2019ajg}), but there will be other multi-center BPS solutions that only exist for the new values of the moduli.  
\end{itemize} 

 Hence, bubbling solutions display the properties of fortuitous states, not only when changing $N$ but also when changing moduli. From this perspective one can say that their ``lifting'' when deforming the theory by changing moduli is a feature, and not a bug!

Unfortunately we do not know the CFT duals of multi-bubble solutions, and so we cannot make a more precise connection with fortuitous states. On the other hand, it would be interesting if one could use the recent study of  fortuitous states in the D1-D5 CFT \cite{Chang:2025rqy}  to identify features of the supergravity solutions describing these states.

We also expect to see fortuity in the supergravity solutions that describe fractionated branes in ten and eleven dimensions (see below).  Brane-fractionation is an intrinsically discontinuous process in both the CFT and in supergravity.  Indeed, breaking up branes into separate clusters often involves a geometric transition in supergravity and this would produce multi-centered solutions akin to those in five dimensions.  We therefore suspect that supergravity will exhibit a very complex combination of fortuity and monotonicity, especially in solutions that probe twisted sectors of the CFT.

}

\section{Supermazes and Supergravity in ten and eleven dimensions}
\label{sec:10-11d}

\subsection{The larger picture: fractionation and themelia}
\label{ss:themelia}

In the vast majority of descriptions of black-hole microstructure, one starts with two species of branes whose numbers are $N_1$ and $N_2$.  Such a system is \nBPS{4}, with eight supersymmetries, and we will refer to this as the {\it substrate.}  To this one can then add $N_P$ momentum excitations perhaps carried by shape modes, densities or fluxes.  This is  the \nBPS{8} solution (with four supersymmetries) that can correspond to a macroscopic black hole, and the details of partitioning the momentum modes into the structure of the substrate will then provide most of  the microstructure.  The goal of the microstate geometry programme is to capture this encoding of momentum modes into the substrate in as accurate a manner as semi-classical analysis will allow.

Brane fractionation {\it within the substrate} is fundamental to the holographic description of twisted sectors of the CFT.   There are two closely related ways in which this can happen:   (i) The intersecting branes of can split and join, to create an object with a much larger world-volume, whose  length that can be up to $N_1 N_2$ times the typical compactification scale, $L_c$ , and thus the energy gap can be as low as  $(L_c N_1 N_2)^{-1}$. Alternatively, or in combination with (i), we have (ii)  The pieces of a multiply-intersecting set of branes can separate from one another so that each individual intersection is described by a separate set of moduli, and there can be  $\sim N_1 N_2$ such moduli.    The black-hole entropy, $S \sim \sqrt{N_1 N_2 N_P}$ then emerges, in (i), from partitioning $N_P$ units of momentum into fractionated excitations, whose momentum is quantized in units of $ (N_1N_2 L_c)^{-1}$ \cite{Strominger:1996sh}, or, for (ii),  into excitations distributed among the $N_1 N_2$ moduli whose momentum is quantized in units of $ L_c^{-1}$ \cite{Maldacena:1997de,Dijkgraaf:1996cv}.  There can also be hybrids of these situations in which the moduli in (ii) are permuted as one goes around another compactification circle, making the world volume much larger, and the energy gap much smaller.

To see, and analyze,  fractionation one needs to work with more than one compactified dimension, and so we need to venture into supergravity in yet more dimensions.  

Supergravity cannot reliably describe a single brane, or a single brane intersection, but it can reliably describe small stacks of branes and multi-intersections of such small stacks. For macroscopic horizons, the numbers $N_1, N_2$ are vast and so it is very reasonable to believe that supergravity should be able to capture a very significant degree of brane fractionation and provide a good physical picture of how brane fractionation happens, as well as how microstructure emerges in collective supergravity modes. With this in mind, one must then select the supergravity system that is most amenable to analysis of the fractionation process.

There is another extremely important idea that emerged from a retrospective analysis of earlier work on microstate geometries:  the idea that microstate structure should be described  by, arguably, the {\em fundamental} structures in string theory, now known as {\it Themelia} \cite{Bena:2022wpl,Bena:2022fzf,Bena:2024qed}.  The detailed constructions of {\it smooth} \nBPS{8} bubbled microstate geometries in five dimensions, and the construction of superstrata, actually resulted in configurations that were {\it locally \nBPS{2}}.  That is, these configurations {\it locally possess}  $16$ supersymmetries:  If one ``zooms in'' on the elements of the overall configuration, one finds, at every point, a sub-configuration  that preserves $16$ supersymmetries.  However, those supersymmetries  depend on where one zooms in;  there are only $4$ common supersymmetries  preserved by the complete configuration when taken as a whole.  This is the defining idea of a themelion; it is what makes it a fundamental object in string theory and  this idea lies at the heart of the themelion's smoothness, constructibility  and  amenability to description in supergravity. 

First, a configuration with $16$ supersymmetries can always be dualized to a stack $N$ branes of a single species.  This not only means that it is a (locally) fundamental object, but also implies that, locally, it can only account for a contribution of $\log(N)$ to the entropy.  Thus the vast majority of its entropy (usually some power of $N$) must appear through classical moduli, like shape modes and densities, and so most of its entropy must be visible within supergravity. Secondly, a configuration with $16$ supersymmetries can always be dualized to a stack of coincident Kaluza Klein Monopoles  and so there is always a duality frame in which it is, at least locally, smooth\footnote{As card-carrying  string theorists we accept $A_N$ orbifold singularities as smooth.}.

Finally we believe (with only circumstantial evidence) that themelia based on momentum carrying fluxes will lead to linear systems of equations for the momentum waves and their fluxes, and hence to a completely characterizable phase space of the BPS excitations that are responsible for the entropy. (We will discuss this further below.)

Thus,  the first efforts in describing brane fractionation in supergravity focussed on momentum carrying themelia, and on eleven-dimensional supergravity because the fields are simple, and much of the essential ground work on the substrates had been explored for intersections of M2 and M5 branes.   

\subsection{Intersecting M2 and M5 branes}
\label{ss:M2M5}

The first task is to  describe the two-charge substrate and the intersecting brane configurations  that result from their (partial) fractionation.  It is also critically important to avoid smearing  the brane intersections because it is precisely this kind of ``ensemble averaging'' that can lead to  singular or degenerate corners of the moduli space and perhaps even create horizons as a result of that averaging \cite{Bena:2022sge}.

The simplest \nBPS{4} brane intersections in M-theory are simply  $N_1$  M2's sharing a common spatial direction (and, of course, time)  with $N_5$  M5's.  The M2 branes and M5 branes can also split into separate clumps and M2 branes can also end on M5 branes. This gives rise to a source in the M5 world-volume fluxes; these fluxes can also have ``sinks'', corresponding to another stack of M2's emerging from the M5's at some other point.  This is the fractionated brane system that underpins the (super)-Maze \cite{Bena:2022wpl,Bena:2024qed,Bena:2024dre}. 

We greatly simplify our task by working with a ``local model'' of brane intersections:  we  decompactify the four-manifold wrapped by the M5 branes and orthogonal to the M2 branes.   Specifically, we will take the M2 branes to lie along the  $(x^0, x^1, x^2) =(t, y, z)$ directions, and  have the $(x^0, x^1) =(t, y)$ directions in common with the M5-branes,  which will extend along the $(x^0, x^1,x^3,x^4,x^5,x^6)=(t,y, \vec u)$ directions. The spatial directions transverse to the branes will be denoted by $(x^7,x^8,x^9,x^{10})= \vec v$.  

The supersymmetries of this brane system satisfy the projection conditions:
\begin{equation}
  \Gamma^{012} \, \varepsilon  ~=~ - \varepsilon \,,   \qquad  \Gamma^{013456} \, \varepsilon  ~=~\varepsilon \,.
 \label{projs1}
\end{equation}
The indices on the gamma matrices are frame indices taken along the  directions of the M2's and M5's. 
Recalling that in eleven dimensions one has $ \Gamma^{0123456789\,10} =  \oneone$, one sees that (\ref{projs1})  implies
\begin{equation}
 \Gamma^{01789\,10} \, \varepsilon  ~=~-\varepsilon \,,
 \label{projs2}
\end{equation}
and hence one can add another set of M5 branes along the directions $01789\,10$ without breaking supersymmetry any further.  We will denote this second possible set of branes by M5', but we will not actively include sources for such branes.

The \nBPS{4} solutions of interest have an eleven-dimensional metric of the form:
\begin{equation}
\begin{aligned}
ds_{11}^2 ~=~  e^{2  A_0}\, \Big[ -dt^2 &~+~ dy^2 ~+~ e^{-3  A_0} \, (-\partial_z w )^{-\frac{1}{2}}\, d \vec u \cdot d \vec u  ~+~ e^{-3  A_0} \, (-\partial_z w )^{\frac{1}{2}}\, d \vec v \cdot d \vec v \,    \\
  &  ~+~  (-\partial_z w ) \, \big( dz ~+~(\partial_z w )^{-1}\,   (\vec \nabla_{\vec u} \, w )  \cdot  d \vec u \big)^2  \Big]\,.
\end{aligned}
 \label{11metric}
\end{equation}
This metric is conformally flat along the common M2-M5 directions, $(t,y) \in \IR^{(1,1)}$, and also along the  M5  directions, parameterized by  $\vec u \in \IR^4$,  and on a  transverse $\IR^4$ parameterized by $\vec v \in \IR^4$.    The metric involves a non-trivial fibration of the ``M-theory direction,'' $z$, over this internal  $\IR^4$.    The  constraints on, and relationships between, the functions $A_0(\vec u, \vec v, z)$ and $w(\vec u, \vec v, z)$ will be discussed below, and, for obvious reasons, we require $\partial_z w <0$.

We  use the following set of frames to define the supersymmetries in (\ref{projs1}):
\begin{equation}
\begin{aligned}
e^0 ~=~  & e^{A_0}\, dt \,, \qquad e^1~=~  e^{A_0}\, dy \,,  \qquad e^2 ~=~   e^{A_0} (-\partial_z w )^{\frac{1}{2}} \, \Big( dz ~+~(\partial_z w )^{-1}\,  \big (\vec \nabla_{\vec u} \, w \big)  \cdot  d \vec u \Big) \,, \\
e^{i+2}~=~  & e^{- \frac{1}{2} A_0} \, (-\partial_z w )^{-\frac{1}{4}}\,  du_i \,, \qquad e^{i+6} ~=~  e^{- \frac{1}{2} A_0} \, (-\partial_z w )^{\frac{1}{4}}\,  dv_i\,,   \qquad {i = 1,2,3,4}   \,.
\end{aligned}
 \label{11frames}
\end{equation}
The three-form vector potential is given by:
\begin{equation}
C^{(3)} ~=~   - e^0 \wedge e^1 \wedge e^2 ~+~ \frac{1}{3!}\, \epsilon_{ijk\ell} \,  \big((\partial_z w )^{-1}\, (\partial_{u_\ell} w) \,  du^i \wedge du^j \wedge du^k ~-~ (\partial_{v_\ell} w)  \, dv^i \wedge dv^j \wedge dv^k  \big)  \,,
 \label{C3gen}
\end{equation}
where $\epsilon_{ijk\ell}$ is the $\epsilon$-symbol on $\IR^4$. 

Define the Laplacians on each $\IR^4$ via:
\begin{equation}
{\cal L}_{\vec u} ~\equiv~  \nabla_{\vec u} \cdot \nabla_{\vec u} \,, \qquad {\cal L}_{\vec v} ~\equiv~  \nabla_{\vec v} \cdot \nabla_{\vec v} \,.
 \label{Laps}
\end{equation}
and suppose that one can find a solution to the generalized Monge-Amp\`ere equation (which we call the ``maze equation''):
\begin{equation}
 {\cal L}_{\vec v} G_0 ~=~  ({\cal L}_{\vec  u} G_0)\,(\partial_z^2 G_0) ~-~ ( \nabla_{\vec u} \partial_z G_0)\cdot  (  \nabla_{\vec u} \partial_z G_0)\,.
 \label{maze-eq}
\end{equation}
The function, $G_0$ provides a pre-potential for all the metric and flux functions:
\begin{equation}
w ~\equiv~ \partial_z G_0  \,, \qquad    e^{-3  A_0} \, (-\partial_z w )^{\frac{1}{2}} ~=~ {\cal L}_{\vec v}  G_0 \,.
 \label{solfns}
\end{equation}
More details can be found in \cite{Bena:2024qed,Bena:2024dre}\,.

While this non-linear equation is daunting, it has been shown that solutions exist, and can be constructed in an iterative expansion that involves solving linear Poisson equations \cite{Lunin:2007mj,Lunin:2008tf}.  Moreover, by imposing spherical symmetry and taking a near-brane limit, one can reduce that maze function to two variables, and the maze equation can be re-cast as a linear system \cite{Lunin:2007ab,DHoker:2008lup,DHoker:2008rje,DHoker:2008wvd,DHoker:2009lky, DHoker:2009wlx,Bobev:2013yra,Bachas:2013vza,Bena:2023rzm} and solved explicitly.  Thus the requisite solutions exist, and can be constructed in explicit detail if one zooms in on each and every intersection.   We will therefore take this solution as  ``given,'' and assume that we have some form of intersecting M2-M5 substrate on which we will erect momentum modes. 

At this point it is worth noting that a very similar thing was encountered in five and six dimensions.  The ``zeroth'' step was to start from an ambi-polar, hyper-K\"ahler metric.  While we typically opted to use a simple hyper-K\"ahler metric, like the Gibbons-Hawking metric,  the general problem requires one to solve a Monge-Amp\`ere equation to determine the general Ricci-flat, K\"ahler metric.  Thus, even in five and six dimensions, the substrate is governed by a non-trivial, non-linear equation.   The difference in eleven (or ten) dimensions is that the only way to reduce the number of variables sufficiently to simplify the non-linear system into an equivalent linear system is to zoom in on intersections.  Thus the only real difference is that here there do not seem to be simple solutions to the maze equation that are asymptotically flat. 

The mathematical ``themelion magic'' happens at the next level: the creation of a three-charge \nBPS{8} configuration by adding momentum to two-charge  \nBPS{4} substrates. To achieve this, the  momentum waves need to be carried by fluxes.  In IIA or IIB supergravity, these waves reflect a fluctuating density of open strings stretched between branes, and in M-theory this is a wave motion in additional fluxes carried by the M2-M5 system.  This also produces a fluctuating back-reaction in the densities and shapes of the original brane substrate.   In five and six dimensions, all these fluxes and momentum waves are governed by a {\it linear system}, like  (\ref{BPSlayer1}) and (\ref{BPSlayer2}), {\it in the background created by the substrate.}  Our hope was that this would carry through in general and, in \cite{Bena:2024qed}, we referred to this as the {\it extended themelion conjecture,} which posits that the linear systems are a feature of the ``glue'' that welds the momentum to the branes  creating an \nBPS{8} solution  that has sixteen local supersymmetries.   One of the primary results of  \cite{Bena:2024qed} was to show that, at least for a simple class of momentum waves, this was indeed true. 

The picture that seems to be emerging from this early work on mazes, and  their IIB analogues, is that, while the ``degree of difficulty'' is much higher, all the remarkable results that gave us five and six-dimensional microstate geometries will generalize to ten and eleven dimensions, and to the description of momentum waves localizing on fractionated branes.

That is, we know that there are configurations of intersecting and fractionated branes that involve sufficiently many branes for the supergravity approximation to be valid.  We know the form of the most generic supergravity solutions that can be sourced by these configurations and have strong arguments that these solutions exist.  Thus we can build \nBPS{4}  substrates in supergravity, even though we do not know the explicit functions in the background.  To this we need to add momentum waves and fluxes, and we have preliminary indications that the equations governing these excitations are linear.  This means that we can characterize the phase space and count momentum states.   The ``icing on the cake'' here is that if one zooms in on the brane intersections, where one expects the momentum waves to localize, one can construct \nBPS{4}  the substrate explicitly and the entire construction of the \nBPS{8}  momentum waves can, in principle, be completed.  So far we have only taken the first few steps in realizing this part of the microstate geometry programme, but those first steps have shown a clear, but challenging, path to reaching the final goal of the entire programme: providing a complete semi-classical description of black-hole microstructure.

  
\section{Final comments}
\label{sec:Conclusions}

As we approach the fiftieth anniversary of supergravity, we should, perhaps, ask ourselves how this massive and enduring enterprise might emerge as a crucially important theory of Nature.   The best answer we can give is through black-hole physics.

The information problem tells us that we need a more complete theory of gravity that can describe quantum aspects of black holes.    The small corrections theorem  \cite{Mathur:2009hf} makes a compelling case that the quantum aspects of black holes must give rise to new physics at the horizon scale of the black hole, and not simply when the curvatures hits the Planck scale, $\ell_P$.  The essential point is that the staggeringly large number, $e^S$, of microstates of a black hole create large-scale collective effects, at distances $\sim  S^\alpha \ell_P$  (for some $\alpha$), that extends out to the horizon scale.  Indeed, in the quantum theory, this is what must define the horizon.  

Given this more recent perspective on black holes and the difficulty of studying such a vast quantum object, one might hope to find an effective field theory that can capture the semi-classical behavior of a black hole.  The obvious zeroth-order answer is, of course, General Relativity, which provides a phenomenally accurate classical description of black holes, but also leads to event horizons and the information problem.  This puts GR in its proper context: it should be viewed as an effective theory and the appearance of an event horizon should be seen as a  failure of that effective field theory: it simply does not have the degrees of freedom to resolve the horizon-scale physics. Indeed, the appearance of a horizon in a theory of gravity becomes the symbol of a failure of the effective field theory: it cannot  describe the microstructure and so it produces a geometric object that reflects an ensemble average over all the details it cannot describe.  The obvious next question is: can we do better, and are there better, more accurate effective field theories?  The answer is a definite yes on both counts.

Before giving our preferred answer to these questions, it is worth reviewing what an acceptable answer must include.  The fundamental theory of black holes  must be able to capture and describe black-hole microstructure and make it accessible to distant observers. This microstructure must be accessible at  the horizon scale of the black hole and encode information on the analogue of outgoing Hawking quanta.   Given such  fundamental requirements, any effective field theory that replaces, or improves on, General Relativity should
\begin{itemize}
\item Provide a support mechanism for horizon-scale structure that prevents the formation of a horizon and the collapse to a singularity. The uniqueness and singularity theorems tell us that using  ``normal  matter'' coupled to GR simply does not work.  

\item Match GR classically, and show how the horizon scale emerges as the break-down scale of GR.   Since the size of the black-hole horizon grows when gravity becomes stronger, this property is, once again, unachievable  with ``normal matter,''  which shrinks when gravity becomes stronger. 

\item   Provide  precise details of how the effective field theory samples the vast phase space of all the black-hole microstates. In this sense, GR is an immensely blunt instrument:  black-hole uniqueness is a statement that GR  fails to resolve any of the microstructure.  It is also not good enough to simply avoid making a horizon but only exhibit a handful of microstates (as happens with boson stars and other black-hole mimickers).  The fundamental theory, whatever it is, must describe all the microstructure, and a ``better'' effective field theory must characterize how it is sampling that vast phase space of microstructure. 

\item Reveal the dynamics of how the horizon-scale structure emerges during gravitational collapse. This property is again highly unusual. Consider a perfectly spherical, collapsing shell of dust whose mass is equal to that of one of the largest black hole in the known Universe, TON618.  This shell will form an event horizon roughly when the its surface gravitational field is the same as the surface gravity on the Earth. At this scale one does not expect any quantum-gravitational effects to be important and to hinder the formation of an event horizon.  The small corrections theorem tells us that, despite the very small scale of the curvature at the horizon,  and despite the perfect spherical symmetry of the dust shell,  quantum effects will dominate over the classical GR evolution and some form of quantum phase transition must happen at the horizon scale.  This quantum phase transition is driven by the vast, $e^{S}$, number of microstate creating a transition amplitude of $\cO(1)$.  A good effective field theory should give some insight into this tunneling process. 
\end{itemize}

From our perspective,  the only viable  quantum theory of gravity is string theory.  What we have described above are the crucial ingredients of the fuzzball programme, and the better effective field theory is supergravity and its exemplification in the microstate geometry programme. (See, for example, \cite{Bena:2022ldq,Bena:2022rna}).  

To expand upon this a little, we note that string theory has already given us a great deal of new insight into how quantum black holes must behave, and has found ways to count black-hole microstructure for supersymmetric black holes in various weak-coupling, non-back-reacted limits. (See, for example, \cite{Sen:1995in, Strominger:1996sh,Maldacena:1997de,Dijkgraaf:1996cv}.)  A better effective field theory must therefore build upon these successes and describe this microstructure at strong coupling and with full back-reaction, and especially with gravitational back-reaction.   This is what has driven the microstate geometry programme, but one should think about what this actually implies more broadly.

Black-hole microstructure must emerge at the horizon scale, and this is a vastly larger scale than the Planck, or string, scale. Therefore, if one is working within string theory, describing any horizon-scale structure can only be done using the massless (long-range) sector. Which means it can only be done using the massless supergravity theories that emerge from string theory.  Supergravity is therefore, of necessity, the repository  of all possible ``better effective field theories.''  The story of microstate geometries is simply an exegesis of this observation.

Five-dimensional supergravities are better effective theories than General Relativity because they show how non-trivial topology and fluxes can provide the essential support mechanism for horizon-scale microstructure \cite{Gibbons:2013tqa}.  Because of the ``smearing'' over the compactification manifold, these theories provide a rather limited class of black-hole microstructure, but do reveal some of the moduli and ground-state structure.  

Six-dimensional supergravities improve on this greatly by incorporating microstructure that can carry  arbitrarily-shaped momentum waves.  The holographic dictionary is thoroughly developed and shows that these geometries are dual to the supergraviton gas states that contributes to the microstate counting in  \cite{Strominger:1996sh}.  The number of microstates visible in six dimensions is vast, but parametrically short of the black-hole entropy.  The problem is that one is still averaging over compactification dimensions and,   in particular, smearing out brane fractionation.
 
The microstate geometry programme in ten and eleven dimensions is  very much under development, but greatly benefiting from the insights gained in five and six dimensions. It can see brane fractionation, which is the origin of the majority of black-hole microstates.  There is a clear semi-classical picture of how this is an effective field theory of precisely the perturbative, weak coupling states that went into the counting of \cite{Sen:1995in, Strominger:1996sh,Maldacena:1997de,Dijkgraaf:1996cv}.  Thus ten and eleven-dimensional supergravities have the potential to provide a complete picture of black-hole microstructure in its coherent, semi-classical limits.

It is also important to note that in other contexts (like quark-gluon plasmas) supergravity has been shown to provide a very effective description of the collective effects of strongly coupled quantum systems.  It is therefore no surprise that supergravity  can provide much better descriptions of quantum black holes  than General Relativity.

Supergravity has come a very long way in fifty years, and it has a very promising future in providing not only better effective descriptions of black-hole physics, but also providing backgrounds that can be used as ``laboratories'' for studying black-hole microstructure.  As supergravity is about to turn 50, it is significant that microstate geometries are celebrating their 20$^{\rm th}$ anniversary  \cite{Bena:2005va,Berglund:2005vb}. We hope that microstate geometries will share the longevity of supergravity, and maybe their ultimate fates in physics are deeply linked through the description of black holes and in the resolution of the information paradox.

\vspace{.3cm}\noindent {\bf Acknowledgements:} \\
We would like to thank our many colleagues, collaborators and students  for their inspiration, insight and hard work over the last two decades. 
The work of IB and NPW was supported in part by the ERC Grant 787320 - QBH Structure and by the DOE grant DE-SC0011687.


\begin{adjustwidth}{-1mm}{-1mm} 

\bibliographystyle{utphys}      

\bibliography{references1}       

\providecommand{\href}[2]{#2}\begingroup\raggedright\begin{thebibliography}{100}

\bibitem{Penrose:1964wq}
R.~Penrose, ``{Gravitational collapse and space-time singularities},''
  \href{http://dx.doi.org/10.1103/PhysRevLett.14.57}{{\em Phys. Rev. Lett.}
  {\bfseries 14} (1965) 57--59}.

\bibitem{Hawking:1976ra}
S.~W. Hawking, ``{Breakdown of Predictability in Gravitational Collapse},''
\href{http://dx.doi.org/10.1103/PhysRevD.14.2460}{{\em Phys. Rev.} {\bfseries
  D14} (1976) 2460--2473}.

\bibitem{Hawking:1982dj}
S.~W. Hawking, ``{The Unpredictability of Quantum Gravity},''
\href{http://dx.doi.org/10.1007/BF01206031}{{\em Commun. Math. Phys.}
  {\bfseries 87} (1982) 395}.

\bibitem{Mathur:2009hf}
S.~D. Mathur, ``{The information paradox: A pedagogical introduction},''
  \href{http://dx.doi.org/10.1088/0264-9381/26/22/224001}{{\em Class. Quant.
  Grav.} {\bfseries 26} (2009) 224001},
\href{http://arxiv.org/abs/0909.1038}{{\ttfamily arXiv:0909.1038 [hep-th]}}.

\bibitem{Mathur:2008kg}
S.~D. Mathur, ``{Tunneling into fuzzball states},''
  \href{http://dx.doi.org/10.1007/s10714-009-0837-3}{{\em Gen. Rel. Grav.}
  {\bfseries 42} (2010) 113--118},
\href{http://arxiv.org/abs/0805.3716}{{\ttfamily arXiv:0805.3716 [hep-th]}}.

\bibitem{Bena:2022ldq}
I.~Bena, E.~J. Martinec, S.~D. Mathur, and N.~P. Warner, ``{Snowmass White
  Paper: Micro- and Macro-Structure of Black Holes},''
  \href{http://arxiv.org/abs/2203.04981}{{\ttfamily arXiv:2203.04981
  [hep-th]}}.

\bibitem{Bena:2022rna}
I.~Bena, E.~J. Martinec, S.~D. Mathur, and N.~P. Warner, ``{Fuzzballs and
  Microstate Geometries: Black-Hole Structure in String Theory},''
  \href{http://arxiv.org/abs/2204.13113}{{\ttfamily arXiv:2204.13113
  [hep-th]}}.

\bibitem{Strominger:1996sh}
A.~Strominger and C.~Vafa, ``{Microscopic Origin of the Bekenstein-Hawking
  Entropy},'' \href{http://dx.doi.org/10.1016/0370-2693(96)00345-0}{{\em Phys.
  Lett.} {\bfseries B379} (1996) 99--104},
\href{http://arxiv.org/abs/hep-th/9601029}{{\ttfamily arXiv:hep-th/9601029}}.

\bibitem{Gauntlett:2002nw}
J.~P. Gauntlett, J.~B. Gutowski, C.~M. Hull, S.~Pakis, and H.~S. Reall, ``{All
  supersymmetric solutions of minimal supergravity in five- dimensions},''
  \href{http://dx.doi.org/10.1088/0264-9381/20/21/005}{{\em Class.Quant.Grav.}
  {\bfseries 20} (2003) 4587--4634},
\href{http://arxiv.org/abs/hep-th/0209114}{{\ttfamily arXiv:hep-th/0209114
  [hep-th]}}.

\bibitem{Bena:2005va}
I.~Bena and N.~P. Warner, ``{Bubbling supertubes and foaming black holes},''
  \href{http://dx.doi.org/10.1103/PhysRevD.74.066001}{{\em Phys. Rev.}
  {\bfseries D74} (2006) 066001},
\href{http://arxiv.org/abs/hep-th/0505166}{{\ttfamily arXiv:hep-th/0505166}}.

\bibitem{Berglund:2005vb}
P.~Berglund, E.~G. Gimon, and T.~S. Levi, ``{Supergravity microstates for BPS
  black holes and black rings},''
  \href{http://dx.doi.org/10.1088/1126-6708/2006/06/007}{{\em JHEP} {\bfseries
  0606} (2006) 007},
\href{http://arxiv.org/abs/hep-th/0505167}{{\ttfamily arXiv:hep-th/0505167
  [hep-th]}}.

\bibitem{Bena:2007kg}
I.~Bena and N.~P. Warner, ``{Black holes, black rings and their microstates},''
  \href{http://dx.doi.org/10.1007/978-3-540-79523-0}{{\em Lect. Notes Phys.}
  {\bfseries 755} (2008) 1--92},
\href{http://arxiv.org/abs/hep-th/0701216}{{\ttfamily arXiv:hep-th/0701216}}.

\bibitem{Bena:2009fi}
I.~Bena, S.~Giusto, C.~Ruef, and N.~P. Warner, ``{Supergravity Solutions from
  Floating Branes},'' \href{http://dx.doi.org/10.1007/JHEP03(2010)047}{{\em
  JHEP} {\bfseries 03} (2010) 047},
\href{http://arxiv.org/abs/0910.1860}{{\ttfamily arXiv:0910.1860 [hep-th]}}.

\bibitem{Bena:2009qv}
I.~Bena, S.~Giusto, C.~Ruef, and N.~P. Warner, ``{A (Running) Bolt for New
  Reasons},'' \href{http://dx.doi.org/10.1088/1126-6708/2009/11/089}{{\em JHEP}
  {\bfseries 11} (2009) 089},
\href{http://arxiv.org/abs/0909.2559}{{\ttfamily arXiv:0909.2559 [hep-th]}}.

\bibitem{Bobev:2009kn}
N.~Bobev and C.~Ruef, ``{The Nuts and Bolts of Einstein-Maxwell Solutions},''
  \href{http://dx.doi.org/10.1007/JHEP01(2010)124}{{\em JHEP} {\bfseries 01}
  (2010) 124}, \href{http://arxiv.org/abs/0912.0010}{{\ttfamily arXiv:0912.0010
  [hep-th]}}.

\bibitem{Bena:2004de}
I.~Bena and N.~P. Warner, ``{One ring to rule them all ... and in the darkness
  bind them?},'' {\em Adv. Theor. Math. Phys.} {\bfseries 9} (2005) 667--701,
\href{http://arxiv.org/abs/hep-th/0408106}{{\ttfamily arXiv:hep-th/0408106}}.

\bibitem{Bena:2006kb}
I.~Bena, C.-W. Wang, and N.~P. Warner, ``{Mergers and Typical Black Hole
  Microstates},'' \href{http://dx.doi.org/10.1088/1126-6708/2006/11/042}{{\em
  JHEP} {\bfseries 11} (2006) 042},
\href{http://arxiv.org/abs/hep-th/0608217}{{\ttfamily arXiv:hep-th/0608217}}.

\bibitem{Bena:2007qc}
I.~Bena, C.-W. Wang, and N.~P. Warner, ``{Plumbing the Abyss: Black Ring
  Microstates},'' \href{http://dx.doi.org/10.1088/1126-6708/2008/07/019}{{\em
  JHEP} {\bfseries 07} (2008) 019},
\href{http://arxiv.org/abs/0706.3786}{{\ttfamily arXiv:0706.3786 [hep-th]}}.

\bibitem{Giusto:2004kj}
S.~Giusto and S.~D. Mathur, ``{Geometry of D1-D5-P bound states},''
  \href{http://dx.doi.org/10.1016/j.nuclphysb.2005.09.037}{{\em Nucl. Phys.}
  {\bfseries B729} (2005) 203--220},
\href{http://arxiv.org/abs/hep-th/0409067}{{\ttfamily arXiv:hep-th/0409067}}.

\bibitem{Bena:2006is}
I.~Bena, C.-W. Wang, and N.~P. Warner, ``{The foaming three-charge black
  hole},'' \href{http://dx.doi.org/10.1103/PhysRevD.75.124026}{{\em Phys. Rev.}
  {\bfseries D75} (2007) 124026},
\href{http://arxiv.org/abs/hep-th/0604110}{{\ttfamily arXiv:hep-th/0604110}}.

\bibitem{Bena:2008wt}
I.~Bena, N.~Bobev, and N.~P. Warner, ``{Spectral Flow, and the Spectrum of
  Multi-Center Solutions},''
  \href{http://dx.doi.org/10.1103/PhysRevD.77.125025}{{\em Phys. Rev.}
  {\bfseries D77} (2008) 125025},
\href{http://arxiv.org/abs/0803.1203}{{\ttfamily arXiv:0803.1203 [hep-th]}}.

\bibitem{Bena:2010gg}
I.~Bena, N.~Bobev, S.~Giusto, C.~Ruef, and N.~P. Warner, ``{An
  Infinite-Dimensional Family of Black-Hole Microstate Geometries},''
  \href{http://dx.doi.org/10.1007/JHEP03(2011)022,
  10.1007/JHEP04(2011)059}{{\em JHEP} {\bfseries 1103} (2011) 022},
  \href{http://arxiv.org/abs/1006.3497}{{\ttfamily arXiv:1006.3497 [hep-th]}}.

\bibitem{Bena:2011dd}
I.~Bena, S.~Giusto, M.~Shigemori, and N.~P. Warner, ``{Supersymmetric Solutions
  in Six Dimensions: A Linear Structure},''
  \href{http://dx.doi.org/10.1007/JHEP03(2012)084}{{\em JHEP} {\bfseries 1203}
  (2012) 084},
\href{http://arxiv.org/abs/1110.2781}{{\ttfamily arXiv:1110.2781 [hep-th]}}.

\bibitem{Bena:2013dka}
I.~Bena and N.~P. Warner, ``{Resolving the Structure of Black Holes:
  Philosophizing with a Hammer},''
\href{http://arxiv.org/abs/1311.4538}{{\ttfamily arXiv:1311.4538 [hep-th]}}.

\bibitem{Bianchi:2016bgx}
M.~Bianchi, J.~F. Morales, and L.~Pieri, ``{Stringy origin of 4d black hole
  microstates},'' \href{http://dx.doi.org/10.1007/JHEP06(2016)003}{{\em JHEP}
  {\bfseries 06} (2016) 003},
\href{http://arxiv.org/abs/1603.05169}{{\ttfamily arXiv:1603.05169 [hep-th]}}.

\bibitem{Bianchi:2017bxl}
M.~Bianchi, J.~F. Morales, L.~Pieri, and N.~Zinnato, ``{More on microstate
  geometries of 4d black holes},''
  \href{http://dx.doi.org/10.1007/JHEP05(2017)147}{{\em JHEP} {\bfseries 05}
  (2017) 147}, \href{http://arxiv.org/abs/1701.05520}{{\ttfamily
  arXiv:1701.05520 [hep-th]}}.

\bibitem{Heidmann:2017cxt}
P.~Heidmann, ``{Four-center bubbled BPS solutions with a Gibbons-Hawking
  base},'' \href{http://dx.doi.org/10.1007/JHEP10(2017)009}{{\em JHEP}
  {\bfseries 10} (2017) 009},
\href{http://arxiv.org/abs/1703.10095}{{\ttfamily arXiv:1703.10095 [hep-th]}}.

\bibitem{Bena:2017fvm}
I.~Bena, P.~Heidmann, and P.~F. Ramirez, ``{A systematic construction of
  microstate geometries with low angular momentum},''
  \href{http://dx.doi.org/10.1007/JHEP10(2017)217}{{\em JHEP} {\bfseries 10}
  (2017) 217},
\href{http://arxiv.org/abs/1709.02812}{{\ttfamily arXiv:1709.02812 [hep-th]}}.

\bibitem{Avila:2017pwi}
J.~Avila, P.~F. Ramirez, and A.~Ruiperez, ``{One Thousand and One Bubbles},''
  \href{http://dx.doi.org/10.1007/JHEP01(2018)041}{{\em JHEP} {\bfseries 01}
  (2018) 041},
\href{http://arxiv.org/abs/1709.03985}{{\ttfamily arXiv:1709.03985 [hep-th]}}.

\bibitem{Tyukov:2018ypq}
A.~Tyukov, R.~Walker, and N.~P. Warner, ``{The Structure of BPS Equations for
  Ambi-polar Microstate Geometries},''
  \href{http://dx.doi.org/10.1088/1361-6382/aaf133}{{\em Class. Quant. Grav.}
  {\bfseries 36} no.~1, (2019) 015021},
\href{http://arxiv.org/abs/1807.06596}{{\ttfamily arXiv:1807.06596 [hep-th]}}.

\bibitem{Warner:2019jll}
N.~P. Warner, ``{Lectures on Microstate Geometries},''
  \href{http://arxiv.org/abs/1912.13108}{{\ttfamily arXiv:1912.13108
  [hep-th]}}.

\bibitem{Rawash:2022sum}
S.~Rawash and D.~Turton, ``{Evolutionary Algorithms for Multi-Center
  Solutions},'' \href{http://dx.doi.org/10.1002/prop.202300255}{{\em Fortsch.
  Phys.} {\bfseries 72} no.~2, (2024) 2300255},
  \href{http://arxiv.org/abs/2212.08585}{{\ttfamily arXiv:2212.08585
  [hep-th]}}.

\bibitem{Goldstein:2008fq}
K.~Goldstein and S.~Katmadas, ``{Almost BPS black holes},''
  \href{http://dx.doi.org/10.1088/1126-6708/2009/05/058}{{\em JHEP} {\bfseries
  05} (2009) 058},
\href{http://arxiv.org/abs/0812.4183}{{\ttfamily arXiv:0812.4183 [hep-th]}}.

\bibitem{Bena:2009ev}
I.~Bena, G.~Dall'Agata, S.~Giusto, C.~Ruef, and N.~P. Warner, ``{Non-BPS Black
  Rings and Black Holes in Taub-NUT},''
  \href{http://dx.doi.org/10.1088/1126-6708/2009/06/015}{{\em JHEP} {\bfseries
  06} (2009) 015},
\href{http://arxiv.org/abs/0902.4526}{{\ttfamily arXiv:0902.4526 [hep-th]}}.

\bibitem{Bena:2009en}
I.~Bena, S.~Giusto, C.~Ruef, and N.~P. Warner, ``{Multi-Center non-BPS Black
  Holes - the Solution},''
  \href{http://dx.doi.org/10.1088/1126-6708/2009/11/032}{{\em JHEP} {\bfseries
  11} (2009) 032},
\href{http://arxiv.org/abs/0908.2121}{{\ttfamily arXiv:0908.2121 [hep-th]}}.

\bibitem{DallAgata:2010srl}
G.~Dall'Agata, S.~Giusto, and C.~Ruef, ``{U-duality and non-BPS solutions},''
  \href{http://dx.doi.org/10.1007/JHEP02(2011)074}{{\em JHEP} {\bfseries 02}
  (2011) 074},
\href{http://arxiv.org/abs/1012.4803}{{\ttfamily arXiv:1012.4803 [hep-th]}}.

\bibitem{Vasilakis:2011ki}
O.~Vasilakis and N.~P. Warner, ``{Mind the Gap: Supersymmetry Breaking in
  Scaling, Microstate Geometries},''
  \href{http://dx.doi.org/10.1007/JHEP10(2011)006}{{\em JHEP} {\bfseries 1110}
  (2011) 006},
\href{http://arxiv.org/abs/1104.2641}{{\ttfamily arXiv:1104.2641 [hep-th]}}.

\bibitem{Heidmann:2018mtx}
P.~Heidmann, ``{Bubbling the NHEK},''
  \href{http://dx.doi.org/10.1007/JHEP01(2019)108}{{\em JHEP} {\bfseries 01}
  (2019) 108}, \href{http://arxiv.org/abs/1811.08256}{{\ttfamily
  arXiv:1811.08256 [hep-th]}}.

\bibitem{Denef:2000nb}
F.~Denef, ``{Supergravity flows and D-brane stability},''
  \href{http://dx.doi.org/10.1088/1126-6708/2000/08/050}{{\em JHEP} {\bfseries
  0008} (2000) 050},
\href{http://arxiv.org/abs/hep-th/0005049}{{\ttfamily arXiv:hep-th/0005049
  [hep-th]}}.

\bibitem{Denef:2002ru}
F.~Denef, ``{Quantum quivers and Hall / hole halos},''
  \href{http://dx.doi.org/10.1088/1126-6708/2002/10/023}{{\em JHEP} {\bfseries
  0210} (2002) 023},
\href{http://arxiv.org/abs/hep-th/0206072}{{\ttfamily arXiv:hep-th/0206072
  [hep-th]}}.

\bibitem{Bates:2003vx}
B.~Bates and F.~Denef, ``{Exact solutions for supersymmetric stationary black
  hole composites},'' \href{http://dx.doi.org/10.1007/JHEP11(2011)127}{{\em
  JHEP} {\bfseries 1111} (2011) 127},
\href{http://arxiv.org/abs/hep-th/0304094}{{\ttfamily arXiv:hep-th/0304094
  [hep-th]}}.

\bibitem{Balasubramanian:2006gi}
V.~Balasubramanian, E.~G. Gimon, and T.~S. Levi, ``{Four Dimensional Black Hole
  Microstates: From D-branes to Spacetime Foam},''
  \href{http://dx.doi.org/10.1088/1126-6708/2008/01/056}{{\em JHEP} {\bfseries
  0801} (2008) 056},
\href{http://arxiv.org/abs/hep-th/0606118}{{\ttfamily arXiv:hep-th/0606118
  [hep-th]}}.

\bibitem{deBoer:2008zn}
J.~de~Boer, S.~El-Showk, I.~Messamah, and D.~Van~den Bleeken, ``{Quantizing N=2
  Multicenter Solutions},''
  \href{http://dx.doi.org/10.1088/1126-6708/2009/05/002}{{\em JHEP} {\bfseries
  05} (2009) 002},
\href{http://arxiv.org/abs/0807.4556}{{\ttfamily arXiv:0807.4556 [hep-th]}}.

\bibitem{deBoer:2009un}
J.~de~Boer, S.~El-Showk, I.~Messamah, and D.~Van~den Bleeken, ``{A bound on the
  entropy of supergravity?},''
  \href{http://dx.doi.org/10.1007/JHEP02(2010)062}{{\em JHEP} {\bfseries 02}
  (2010) 062},
\href{http://arxiv.org/abs/0906.0011}{{\ttfamily arXiv:0906.0011 [hep-th]}}.

\bibitem{Martinec:2015pfa}
E.~J. Martinec and B.~E. Niehoff, ``{Hair-brane Ideas on the Horizon},''
  \href{http://dx.doi.org/10.1007/JHEP11(2015)195}{{\em JHEP} {\bfseries 11}
  (2015) 195},
\href{http://arxiv.org/abs/1509.00044}{{\ttfamily arXiv:1509.00044 [hep-th]}}.

\bibitem{Li:2021gbg}
Y.~Li, ``{Black holes and the swampland: the deep throat revelations},''
  \href{http://dx.doi.org/10.1007/JHEP06(2021)065}{{\em JHEP} {\bfseries 06}
  (2021) 065}, \href{http://arxiv.org/abs/2102.04480}{{\ttfamily
  arXiv:2102.04480 [hep-th]}}.

\bibitem{Gibbons:2013tqa}
G.~Gibbons and N.~Warner, ``{Global structure of five-dimensional fuzzballs},''
  \href{http://dx.doi.org/10.1088/0264-9381/31/2/025016}{{\em
  Class.Quant.Grav.} {\bfseries 31} (2014) 025016},
\href{http://arxiv.org/abs/1305.0957}{{\ttfamily arXiv:1305.0957 [hep-th]}}.

\bibitem{deLange:2015gca}
P.~de~Lange, D.~R. Mayerson, and B.~Vercnocke, ``{Structure of Six-Dimensional
  Microstate Geometries},''
  \href{http://dx.doi.org/10.1007/JHEP09(2015)075}{{\em JHEP} {\bfseries 09}
  (2015) 075},
\href{http://arxiv.org/abs/1504.07987}{{\ttfamily arXiv:1504.07987 [hep-th]}}.

\bibitem{Bah:2021owp}
I.~Bah and P.~Heidmann, ``{Smooth bubbling geometries without supersymmetry},''
  \href{http://dx.doi.org/10.1007/JHEP09(2021)128}{{\em JHEP} {\bfseries 09}
  (2021) 128}, \href{http://arxiv.org/abs/2106.05118}{{\ttfamily
  arXiv:2106.05118 [hep-th]}}.

\bibitem{Bah:2021rki}
I.~Bah and P.~Heidmann, ``{Bubble bag end: a bubbly resolution of curvature
  singularity},'' \href{http://dx.doi.org/10.1007/JHEP10(2021)165}{{\em JHEP}
  {\bfseries 10} (2021) 165}, \href{http://arxiv.org/abs/2107.13551}{{\ttfamily
  arXiv:2107.13551 [hep-th]}}.

\bibitem{Heidmann:2021cms}
P.~Heidmann, ``{Non-BPS floating branes and bubbling geometries},''
  \href{http://dx.doi.org/10.1007/JHEP02(2022)162}{{\em JHEP} {\bfseries 02}
  (2022) 162}, \href{http://arxiv.org/abs/2112.03279}{{\ttfamily
  arXiv:2112.03279 [hep-th]}}.

\bibitem{Bah:2022yji}
I.~Bah, P.~Heidmann, and P.~Weck, ``{Schwarzschild-like topological
  solitons},'' \href{http://dx.doi.org/10.1007/JHEP08(2022)269}{{\em JHEP}
  {\bfseries 08} (2022) 269}, \href{http://arxiv.org/abs/2203.12625}{{\ttfamily
  arXiv:2203.12625 [hep-th]}}.

\bibitem{Bah:2022pdn}
I.~Bah and P.~Heidmann, ``{Non-BPS bubbling geometries in AdS$_{3}$},''
  \href{http://dx.doi.org/10.1007/JHEP02(2023)133}{{\em JHEP} {\bfseries 02}
  (2023) 133}, \href{http://arxiv.org/abs/2210.06483}{{\ttfamily
  arXiv:2210.06483 [hep-th]}}.

\bibitem{Heidmann:2022zyd}
P.~Heidmann and A.~Houppe, ``{Solitonic excitations in AdS$_{2}$},''
  \href{http://dx.doi.org/10.1007/JHEP07(2023)186}{{\em JHEP} {\bfseries 07}
  (2023) 186}, \href{http://arxiv.org/abs/2212.05065}{{\ttfamily
  arXiv:2212.05065 [hep-th]}}.

\bibitem{Bah:2023ows}
I.~Bah and P.~Heidmann, ``{Geometric resolution of the Schwarzschild
  horizon},'' \href{http://dx.doi.org/10.1103/PhysRevD.109.066014}{{\em Phys.
  Rev. D} {\bfseries 109} no.~6, (2024) 066014},
  \href{http://arxiv.org/abs/2303.10186}{{\ttfamily arXiv:2303.10186
  [hep-th]}}.

\bibitem{Mateos:2001qs}
D.~Mateos and P.~K. Townsend, ``{Supertubes},''
  \href{http://dx.doi.org/10.1103/PhysRevLett.87.011602}{{\em Phys. Rev. Lett.}
  {\bfseries 87} (2001) 011602},
\href{http://arxiv.org/abs/hep-th/0103030}{{\ttfamily arXiv:hep-th/0103030}}.

\bibitem{Emparan:2001ux}
R.~Emparan, D.~Mateos, and P.~K. Townsend, ``{Supergravity supertubes},'' {\em
  JHEP} {\bfseries 07} (2001) 011,
\href{http://arxiv.org/abs/hep-th/0106012}{{\ttfamily arXiv:hep-th/0106012}}.

\bibitem{Lunin:2001fv}
O.~Lunin and S.~D. Mathur, ``{Metric of the multiply wound rotating string},''
  \href{http://dx.doi.org/10.1016/S0550-3213(01)00321-2}{{\em Nucl. Phys.}
  {\bfseries B610} (2001) 49--76},
\href{http://arxiv.org/abs/hep-th/0105136}{{\ttfamily arXiv:hep-th/0105136}}.

\bibitem{Lunin:2002iz}
O.~Lunin, J.~M. Maldacena, and L.~Maoz, ``{Gravity solutions for the D1-D5
  system with angular momentum},''
\href{http://arxiv.org/abs/hep-th/0212210}{{\ttfamily arXiv:hep-th/0212210}}.

\bibitem{Palmer:2004gu}
B.~C. Palmer and D.~Marolf, ``{Counting supertubes},''
  \href{http://dx.doi.org/10.1088/1126-6708/2004/06/028}{{\em JHEP} {\bfseries
  06} (2004) 028},
\href{http://arxiv.org/abs/hep-th/0403025}{{\ttfamily arXiv:hep-th/0403025}}.

\bibitem{Rychkov:2005ji}
V.~S. Rychkov, ``{D1-D5 black hole microstate counting from supergravity},''
  \href{http://dx.doi.org/10.1088/1126-6708/2006/01/063}{{\em JHEP} {\bfseries
  01} (2006) 063},
\href{http://arxiv.org/abs/hep-th/0512053}{{\ttfamily arXiv:hep-th/0512053}}.

\bibitem{Bena:2008nh}
I.~Bena, N.~Bobev, C.~Ruef, and N.~P. Warner, ``{Entropy Enhancement and Black
  Hole Microstates},''
  \href{http://dx.doi.org/10.1103/PhysRevLett.105.231301}{{\em Phys. Rev.
  Lett.} {\bfseries 105} (2010) 231301},
\href{http://arxiv.org/abs/0804.4487}{{\ttfamily arXiv:0804.4487 [hep-th]}}.

\bibitem{Bena:2008dw}
I.~Bena, N.~Bobev, C.~Ruef, and N.~P. Warner, ``{Supertubes in Bubbling
  Backgrounds: Born-Infeld Meets Supergravity},''
  \href{http://dx.doi.org/10.1088/1126-6708/2009/07/106}{{\em JHEP} {\bfseries
  07} (2009) 106},
\href{http://arxiv.org/abs/0812.2942}{{\ttfamily arXiv:0812.2942 [hep-th]}}.

\bibitem{Gutowski:2003rg}
J.~B. Gutowski, D.~Martelli, and H.~S. Reall, ``{All supersymmetric solutions
  of minimal supergravity in six dimensions},''
  \href{http://dx.doi.org/10.1088/0264-9381/20/23/008}{{\em Class. Quant.
  Grav.} {\bfseries 20} (2003) 5049--5078},
\href{http://arxiv.org/abs/hep-th/0306235}{{\ttfamily arXiv:hep-th/0306235}}.

\bibitem{Cariglia:2004kk}
M.~Cariglia and O.~A.~P. Mac~Conamhna, ``{The General form of supersymmetric
  solutions of N=(1,0) U(1) and SU(2) gauged supergravities in
  six-dimensions},'' \href{http://dx.doi.org/10.1088/0264-9381/21/13/006}{{\em
  Class. Quant. Grav.} {\bfseries 21} (2004) 3171--3196},
\href{http://arxiv.org/abs/hep-th/0402055}{{\ttfamily arXiv:hep-th/0402055
  [hep-th]}}.

\bibitem{Giusto:2013rxa}
S.~Giusto, L.~Martucci, M.~Petrini, and R.~Russo, ``{6D microstate geometries
  from 10D structures},''
  \href{http://dx.doi.org/10.1016/j.nuclphysb.2013.08.018}{{\em Nucl.Phys.}
  {\bfseries B876} (2013) 509--555},
\href{http://arxiv.org/abs/1306.1745}{{\ttfamily arXiv:1306.1745 [hep-th]}}.

\bibitem{Bena:2017geu}
I.~Bena, E.~Martinec, D.~Turton, and N.~P. Warner, ``{M-theory Superstrata and
  the MSW String},'' \href{http://dx.doi.org/10.1007/JHEP06(2017)137}{{\em
  JHEP} {\bfseries 06} (2017) 137},
\href{http://arxiv.org/abs/1703.10171}{{\ttfamily arXiv:1703.10171 [hep-th]}}.

\bibitem{Giusto:2011fy}
S.~Giusto, R.~Russo, and D.~Turton, ``{New D1-D5-P geometries from string
  amplitudes},'' \href{http://dx.doi.org/10.1007/JHEP11(2011)062}{{\em JHEP}
  {\bfseries 11} (2011) 062},
\href{http://arxiv.org/abs/1108.6331}{{\ttfamily arXiv:1108.6331 [hep-th]}}.

\bibitem{Bena:2015bea}
I.~Bena, S.~Giusto, R.~Russo, M.~Shigemori, and N.~P. Warner, ``{Habemus
  Superstratum! A constructive proof of the existence of superstrata},''
  \href{http://dx.doi.org/10.1007/JHEP05(2015)110}{{\em JHEP} {\bfseries 05}
  (2015) 110},
\href{http://arxiv.org/abs/1503.01463}{{\ttfamily arXiv:1503.01463 [hep-th]}}.

\bibitem{Bena:2016agb}
I.~Bena, E.~Martinec, D.~Turton, and N.~P. Warner, ``{Momentum Fractionation on
  Superstrata},'' \href{http://dx.doi.org/10.1007/JHEP05(2016)064}{{\em JHEP}
  {\bfseries 05} (2016) 064},
\href{http://arxiv.org/abs/1601.05805}{{\ttfamily arXiv:1601.05805 [hep-th]}}.

\bibitem{Bena:2016ypk}
I.~Bena, S.~Giusto, E.~J. Martinec, R.~Russo, M.~Shigemori, D.~Turton, and
  N.~P. Warner, ``{Smooth horizonless geometries deep inside the black-hole
  regime},'' \href{http://dx.doi.org/10.1103/PhysRevLett.117.201601}{{\em Phys.
  Rev. Lett.} {\bfseries 117} no.~20, (2016) 201601},
\href{http://arxiv.org/abs/1607.03908}{{\ttfamily arXiv:1607.03908 [hep-th]}}.

\bibitem{Bena:2017xbt}
I.~Bena, S.~Giusto, E.~J. Martinec, R.~Russo, M.~Shigemori, D.~Turton, and
  N.~P. Warner, ``{Asymptotically-flat supergravity solutions deep inside the
  black-hole regime},'' \href{http://dx.doi.org/10.1007/JHEP02(2018)014}{{\em
  JHEP} {\bfseries 02} (2018) 014},
\href{http://arxiv.org/abs/1711.10474}{{\ttfamily arXiv:1711.10474 [hep-th]}}.

\bibitem{Bakhshaei:2018vux}
E.~Bakhshaei and A.~Bombini, ``{Three-charge superstrata with internal
  excitations},'' \href{http://dx.doi.org/10.1088/1361-6382/ab01bc}{{\em Class.
  Quant. Grav.} {\bfseries 36} no.~5, (2019) 055001},
  \href{http://arxiv.org/abs/1811.00067}{{\ttfamily arXiv:1811.00067
  [hep-th]}}.

\bibitem{Ceplak:2018pws}
N.~{\v C}eplak, R.~Russo, and M.~Shigemori, ``{Supercharging Superstrata},''
  \href{http://dx.doi.org/10.1007/JHEP03(2019)095}{{\em JHEP} {\bfseries 03}
  (2019) 095}, \href{http://arxiv.org/abs/1812.08761}{{\ttfamily
  arXiv:1812.08761 [hep-th]}}.

\bibitem{Heidmann:2019zws}
P.~Heidmann and N.~P. Warner, ``{Superstratum Symbiosis},''
  \href{http://dx.doi.org/10.1007/JHEP09(2019)059}{{\em JHEP} {\bfseries 09}
  (2019) 059}, \href{http://arxiv.org/abs/1903.07631}{{\ttfamily
  arXiv:1903.07631 [hep-th]}}.

\bibitem{Mayerson:2020tcl}
D.~R. Mayerson, R.~A. Walker, and N.~P. Warner, ``{Microstate Geometries from
  Gauged Supergravity in Three Dimensions},''
  \href{http://dx.doi.org/10.1007/JHEP10(2020)030}{{\em JHEP} {\bfseries 10}
  (2020) 030}, \href{http://arxiv.org/abs/2004.13031}{{\ttfamily
  arXiv:2004.13031 [hep-th]}}.

\bibitem{Ganchev:2021iwy}
B.~Ganchev, A.~Houppe, and N.~P. Warner, ``{New superstrata from
  three-dimensional supergravity},''
  \href{http://dx.doi.org/10.1007/JHEP04(2022)065}{{\em JHEP} {\bfseries 04}
  (2022) 065}, \href{http://arxiv.org/abs/2110.02961}{{\ttfamily
  arXiv:2110.02961 [hep-th]}}.

\bibitem{Heidmann:2019xrd}
P.~Heidmann, D.~R. Mayerson, R.~Walker, and N.~P. Warner, ``{Holomorphic Waves
  of Black Hole Microstructure},''
  \href{http://dx.doi.org/10.1007/JHEP02(2020)192}{{\em JHEP} {\bfseries 02}
  (2020) 192}, \href{http://arxiv.org/abs/1910.10714}{{\ttfamily
  arXiv:1910.10714 [hep-th]}}.

\bibitem{Giusto:2015dfa}
S.~Giusto, E.~Moscato, and R.~Russo, ``{AdS$_{3}$ holography for 1/4 and 1/8
  BPS geometries},'' \href{http://dx.doi.org/10.1007/JHEP11(2015)004}{{\em
  JHEP} {\bfseries 11} (2015) 004},
\href{http://arxiv.org/abs/1507.00945}{{\ttfamily arXiv:1507.00945 [hep-th]}}.

\bibitem{Galliani:2016cai}
A.~Galliani, S.~Giusto, E.~Moscato, and R.~Russo, ``{Correlators at large c
  without information loss},''
  \href{http://dx.doi.org/10.1007/JHEP09(2016)065}{{\em JHEP} {\bfseries 09}
  (2016) 065}, \href{http://arxiv.org/abs/1606.01119}{{\ttfamily
  arXiv:1606.01119 [hep-th]}}.

\bibitem{Bombini:2017sge}
A.~Bombini, A.~Galliani, S.~Giusto, E.~Moscato, and R.~Russo, ``{Unitary
  4-point correlators from classical geometries},''
  \href{http://dx.doi.org/10.1140/epjc/s10052-017-5492-3}{{\em Eur. Phys. J.}
  {\bfseries C78} no.~1, (2018) 8},
\href{http://arxiv.org/abs/1710.06820}{{\ttfamily arXiv:1710.06820 [hep-th]}}.

\bibitem{Bombini:2019vnc}
A.~Bombini and A.~Galliani, ``{AdS$_{3}$ four-point functions from $
  \frac{1}{8} $ -BPS states},''
  \href{http://dx.doi.org/10.1007/JHEP06(2019)044}{{\em JHEP} {\bfseries 06}
  (2019) 044}, \href{http://arxiv.org/abs/1904.02656}{{\ttfamily
  arXiv:1904.02656 [hep-th]}}.

\bibitem{Tian:2019ash}
J.~Tian, J.~Hou, and B.~Chen, ``{Holographic Correlators on Integrable
  Superstrata},'' \href{http://dx.doi.org/10.1016/j.nuclphysb.2019.114766}{{\em
  Nucl. Phys. B} {\bfseries 948} (2019) 114766},
  \href{http://arxiv.org/abs/1904.04532}{{\ttfamily arXiv:1904.04532
  [hep-th]}}.

\bibitem{Tormo:2019yus}
J.~Garcia~i Tormo and M.~Taylor, ``{One point functions for black hole
  microstates},'' \href{http://dx.doi.org/10.1007/s10714-019-2566-6}{{\em Gen.
  Rel. Grav.} {\bfseries 51} no.~7, (2019) 89},
  \href{http://arxiv.org/abs/1904.10200}{{\ttfamily arXiv:1904.10200
  [hep-th]}}.

\bibitem{Giusto:2019qig}
S.~Giusto, S.~Rawash, and D.~Turton, ``{Ads$_{3}$ holography at dimension
  two},'' \href{http://dx.doi.org/10.1007/JHEP07(2019)171}{{\em JHEP}
  {\bfseries 07} (2019) 171}, \href{http://arxiv.org/abs/1904.12880}{{\ttfamily
  arXiv:1904.12880 [hep-th]}}.

\bibitem{Bena:2019azk}
I.~Bena, P.~Heidmann, R.~Monten, and N.~P. Warner, ``{Thermal Decay without
  Information Loss in Horizonless Microstate Geometries},''
\href{http://arxiv.org/abs/1905.05194}{{\ttfamily arXiv:1905.05194 [hep-th]}}.

\bibitem{Giusto:2019pxc}
S.~Giusto, R.~Russo, A.~Tyukov, and C.~Wen, ``{Holographic correlators in
  AdS$_3$ without Witten diagrams},''
  \href{http://dx.doi.org/10.1007/JHEP09(2019)030}{{\em JHEP} {\bfseries 09}
  (2019) 030}, \href{http://arxiv.org/abs/1905.12314}{{\ttfamily
  arXiv:1905.12314 [hep-th]}}.

\bibitem{Hulik:2019pwr}
O.~Hulik, J.~Raeymaekers, and O.~Vasilakis, ``{Information recovery from pure
  state geometries in 3D},''
  \href{http://dx.doi.org/10.1007/JHEP06(2020)119}{{\em JHEP} {\bfseries 06}
  (2020) 119}, \href{http://arxiv.org/abs/1911.12309}{{\ttfamily
  arXiv:1911.12309 [hep-th]}}.

\bibitem{Giusto:2020mup}
S.~Giusto, M.~R.~R. Hughes, and R.~Russo, ``{The Regge limit of AdS$_{3}$
  holographic correlators},''
  \href{http://dx.doi.org/10.1007/JHEP11(2020)018}{{\em JHEP} {\bfseries 11}
  (2020) 018}, \href{http://arxiv.org/abs/2007.12118}{{\ttfamily
  arXiv:2007.12118 [hep-th]}}.

\bibitem{Ceplak:2021wzz}
N.~Ceplak, S.~Giusto, M.~R.~R. Hughes, and R.~Russo, ``{Holographic correlators
  with multi-particle states},''
  \href{http://dx.doi.org/10.1007/JHEP09(2021)204}{{\em JHEP} {\bfseries 09}
  (2021) 204}, \href{http://arxiv.org/abs/2105.04670}{{\ttfamily
  arXiv:2105.04670 [hep-th]}}.

\bibitem{Rawash:2021pik}
S.~Rawash and D.~Turton, ``{Supercharged AdS$_{3}$ Holography},''
  \href{http://dx.doi.org/10.1007/JHEP07(2021)178}{{\em JHEP} {\bfseries 07}
  (2021) 178}, \href{http://arxiv.org/abs/2105.13046}{{\ttfamily
  arXiv:2105.13046 [hep-th]}}.

\bibitem{Ganchev:2021pgs}
B.~Ganchev, A.~Houppe, and N.~P. Warner, ``{Q-balls meet fuzzballs: non-BPS
  microstate geometries},''
  \href{http://dx.doi.org/10.1007/JHEP11(2021)028}{{\em JHEP} {\bfseries 11}
  (2021) 028}, \href{http://arxiv.org/abs/2107.09677}{{\ttfamily
  arXiv:2107.09677 [hep-th]}}.

\bibitem{Ganchev:2021ewa}
B.~Ganchev, S.~Giusto, A.~Houppe, and R.~Russo, ``{$\hbox {AdS}_3$ holography
  for non-BPS geometries},''
  \href{http://dx.doi.org/10.1140/epjc/s10052-022-10133-2}{{\em Eur. Phys. J.
  C} {\bfseries 82} no.~3, (2022) 217},
  \href{http://arxiv.org/abs/2112.03287}{{\ttfamily arXiv:2112.03287
  [hep-th]}}.

\bibitem{Ganchev:2023sth}
B.~Ganchev, S.~Giusto, A.~Houppe, R.~Russo, and N.~P. Warner,
  ``{Microstrata},'' \href{http://dx.doi.org/10.1007/JHEP10(2023)163}{{\em
  JHEP} {\bfseries 10} (2023) 163},
  \href{http://arxiv.org/abs/2307.13021}{{\ttfamily arXiv:2307.13021
  [hep-th]}}.

\bibitem{Houppe:2024hyj}
A.~Houppe, ``{Time-dependent microstrata in AdS$_{3}$},''
  \href{http://dx.doi.org/10.1007/JHEP09(2024)083}{{\em JHEP} {\bfseries 09}
  (2024) 083}, \href{http://arxiv.org/abs/2402.11017}{{\ttfamily
  arXiv:2402.11017 [hep-th]}}.

\bibitem{Bena:2022sge}
I.~Bena, N.~Ceplak, S.~Hampton, Y.~Li, D.~Toulikas, and N.~P. Warner,
  ``{Resolving black-hole microstructure with new momentum carriers},''
  \href{http://dx.doi.org/10.1007/JHEP10(2022)033}{{\em JHEP} {\bfseries 10}
  (2022) 033}, \href{http://arxiv.org/abs/2202.08844}{{\ttfamily
  arXiv:2202.08844 [hep-th]}}.

\bibitem{Martinec:2022okx}
E.~J. Martinec, S.~Massai, and D.~Turton, ``{On the BPS Sector in AdS3/CFT2
  Holography},'' \href{http://dx.doi.org/10.1002/prop.202300015}{{\em Fortsch.
  Phys.} {\bfseries 71} no.~4-5, (2023) 2300015},
  \href{http://arxiv.org/abs/2211.12476}{{\ttfamily arXiv:2211.12476
  [hep-th]}}.

\bibitem{Ceplak:2022wri}
N.~\v{C}eplak, S.~Hampton, and N.~P. Warner, ``{Linearizing the BPS equations
  with vector and tensor multiplets},''
  \href{http://dx.doi.org/10.1007/JHEP03(2023)145}{{\em JHEP} {\bfseries 03}
  (2023) 145}, \href{http://arxiv.org/abs/2204.07170}{{\ttfamily
  arXiv:2204.07170 [hep-th]}}.

\bibitem{Ceplak:2022pep}
N.~\v{C}eplak, ``{Vector Superstrata},''
  \href{http://dx.doi.org/10.1007/JHEP08(2023)047}{{\em JHEP} {\bfseries 08}
  (2023) 047}, \href{http://arxiv.org/abs/2212.06947}{{\ttfamily
  arXiv:2212.06947 [hep-th]}}.

\bibitem{Ceplak:2024dbj}
N.~\v{C}eplak and S.~D. Hampton, ``{Vector superstrata. Part II},''
  \href{http://dx.doi.org/10.1007/JHEP10(2024)011}{{\em JHEP} {\bfseries 10}
  (2024) 011}, \href{http://arxiv.org/abs/2405.05341}{{\ttfamily
  arXiv:2405.05341 [hep-th]}}.

\bibitem{Shigemori:2019orj}
M.~Shigemori, ``{Counting Superstrata},''
\href{http://arxiv.org/abs/1907.03878}{{\ttfamily arXiv:1907.03878 [hep-th]}}.

\bibitem{Mayerson:2020acj}
D.~R. Mayerson and M.~Shigemori, ``{Counting D1-D5-P microstates in
  supergravity},'' \href{http://dx.doi.org/10.21468/SciPostPhys.10.1.018}{{\em
  SciPost Phys.} {\bfseries 10} no.~1, (2021) 018},
  \href{http://arxiv.org/abs/2010.04172}{{\ttfamily arXiv:2010.04172
  [hep-th]}}.

\bibitem{Shigemori:2022gxf}
M.~Shigemori, ``{Superstrata on orbifolded backgrounds},''
  \href{http://dx.doi.org/10.1007/JHEP02(2023)099}{{\em JHEP} {\bfseries 02}
  (2023) 099}, \href{http://arxiv.org/abs/2212.13388}{{\ttfamily
  arXiv:2212.13388 [hep-th]}}.

\bibitem{Chang:2024zqi}
C.-M. Chang and Y.-H. Lin, ``{Holographic covering and the fortuity of black
  holes},'' \href{http://arxiv.org/abs/2402.10129}{{\ttfamily arXiv:2402.10129
  [hep-th]}}.

\bibitem{Chang:2025rqy}
C.-M. Chang, Y.-H. Lin, and H.~Zhang, ``{Fortuity in the D1-D5 system},''
  \href{http://arxiv.org/abs/2501.05448}{{\ttfamily arXiv:2501.05448
  [hep-th]}}.

\bibitem{Bossard:2019ajg}
G.~Bossard and S.~Luest, ``{Microstate geometries at a generic point in moduli
  space},'' \href{http://dx.doi.org/10.1007/s10714-019-2584-4}{{\em Gen. Rel.
  Grav.} {\bfseries 51} no.~9, (2019) 112},
  \href{http://arxiv.org/abs/1905.12012}{{\ttfamily arXiv:1905.12012
  [hep-th]}}.

\bibitem{Maldacena:1997de}
J.~M. Maldacena, A.~Strominger, and E.~Witten, ``{Black hole entropy in
  M-theory},'' {\em JHEP} {\bfseries 12} (1997) 002,
\href{http://arxiv.org/abs/hep-th/9711053}{{\ttfamily arXiv:hep-th/9711053}}.

\bibitem{Dijkgraaf:1996cv}
R.~Dijkgraaf, E.~P. Verlinde, and H.~L. Verlinde, ``{BPS spectrum of the
  five-brane and black hole entropy},''
  \href{http://dx.doi.org/10.1016/S0550-3213(96)00638-4}{{\em Nucl. Phys. B}
  {\bfseries 486} (1997) 77--88},
  \href{http://arxiv.org/abs/hep-th/9603126}{{\ttfamily arXiv:hep-th/9603126}}.

\bibitem{Bena:2022wpl}
I.~Bena, S.~D. Hampton, A.~Houppe, Y.~Li, and D.~Toulikas, ``{The (amazing)
  super-maze},'' \href{http://dx.doi.org/10.1007/JHEP03(2023)237}{{\em JHEP}
  {\bfseries 03} (2023) 237}, \href{http://arxiv.org/abs/2211.14326}{{\ttfamily
  arXiv:2211.14326 [hep-th]}}.

\bibitem{Bena:2022fzf}
I.~Bena, N.~\v{C}eplak, S.~D. Hampton, A.~Houppe, D.~Toulikas, and N.~P.
  Warner, ``{Themelia: the irreducible microstructure of black holes},''
  \href{http://arxiv.org/abs/2212.06158}{{\ttfamily arXiv:2212.06158
  [hep-th]}}.

\bibitem{Bena:2024qed}
I.~Bena, R.~Dulac, A.~Houppe, D.~Toulikas, and N.~P. Warner, ``{Waves on
  Mazes},'' \href{http://arxiv.org/abs/2404.14477}{{\ttfamily arXiv:2404.14477
  [hep-th]}}.

\bibitem{Bena:2024dre}
I.~Bena, S.~Chakraborty, D.~Toulikas, and N.~P. Warner, ``{The M2-M5 Mohawk},''
  \href{http://arxiv.org/abs/2407.01665}{{\ttfamily arXiv:2407.01665
  [hep-th]}}.

\bibitem{Lunin:2007mj}
O.~Lunin, ``{Strings ending on branes from supergravity},''
  \href{http://dx.doi.org/10.1088/1126-6708/2007/09/093}{{\em JHEP} {\bfseries
  09} (2007) 093}, \href{http://arxiv.org/abs/0706.3396}{{\ttfamily
  arXiv:0706.3396 [hep-th]}}.

\bibitem{Lunin:2008tf}
O.~Lunin, ``{Brane webs and 1/4-BPS geometries},''
  \href{http://dx.doi.org/10.1088/1126-6708/2008/09/028}{{\em JHEP} {\bfseries
  0809} (2008) 028},
\href{http://arxiv.org/abs/0802.0735}{{\ttfamily arXiv:0802.0735 [hep-th]}}.

\bibitem{Lunin:2007ab}
O.~Lunin, ``{1/2-BPS states in M theory and defects in the dual CFTs},''
  \href{http://dx.doi.org/10.1088/1126-6708/2007/10/014}{{\em JHEP} {\bfseries
  10} (2007) 014}, \href{http://arxiv.org/abs/0704.3442}{{\ttfamily
  arXiv:0704.3442 [hep-th]}}.

\bibitem{DHoker:2008lup}
E.~D'Hoker, J.~Estes, M.~Gutperle, and D.~Krym, ``{Exact Half-BPS Flux
  Solutions in M-theory. I: Local Solutions},''
  \href{http://dx.doi.org/10.1088/1126-6708/2008/08/028}{{\em JHEP} {\bfseries
  08} (2008) 028}, \href{http://arxiv.org/abs/0806.0605}{{\ttfamily
  arXiv:0806.0605 [hep-th]}}.

\bibitem{DHoker:2008rje}
E.~D'Hoker, J.~Estes, M.~Gutperle, and D.~Krym, ``{Exact Half-BPS Flux
  Solutions in M-theory II: Global solutions asymptotic to AdS(7) x S**4},''
  \href{http://dx.doi.org/10.1088/1126-6708/2008/12/044}{{\em JHEP} {\bfseries
  12} (2008) 044}, \href{http://arxiv.org/abs/0810.4647}{{\ttfamily
  arXiv:0810.4647 [hep-th]}}.

\bibitem{DHoker:2008wvd}
E.~D'Hoker, J.~Estes, M.~Gutperle, D.~Krym, and P.~Sorba, ``{Half-BPS
  supergravity solutions and superalgebras},''
  \href{http://dx.doi.org/10.1088/1126-6708/2008/12/047}{{\em JHEP} {\bfseries
  12} (2008) 047}, \href{http://arxiv.org/abs/0810.1484}{{\ttfamily
  arXiv:0810.1484 [hep-th]}}.

\bibitem{DHoker:2009lky}
E.~D'Hoker, J.~Estes, M.~Gutperle, and D.~Krym, ``{Janus solutions in
  M-theory},'' \href{http://dx.doi.org/10.1088/1126-6708/2009/06/018}{{\em
  JHEP} {\bfseries 06} (2009) 018},
  \href{http://arxiv.org/abs/0904.3313}{{\ttfamily arXiv:0904.3313 [hep-th]}}.

\bibitem{DHoker:2009wlx}
E.~D'Hoker, J.~Estes, M.~Gutperle, and D.~Krym, ``{Exact Half-BPS Flux
  Solutions in M-theory III: Existence and rigidity of global solutions
  asymptotic to AdS(4) x S**7},''
  \href{http://dx.doi.org/10.1088/1126-6708/2009/09/067}{{\em JHEP} {\bfseries
  09} (2009) 067}, \href{http://arxiv.org/abs/0906.0596}{{\ttfamily
  arXiv:0906.0596 [hep-th]}}.

\bibitem{Bobev:2013yra}
N.~Bobev, K.~Pilch, and N.~P. Warner, ``{Supersymmetric Janus Solutions in Four
  Dimensions},'' \href{http://dx.doi.org/10.1007/JHEP06(2014)058}{{\em JHEP}
  {\bfseries 06} (2014) 058}, \href{http://arxiv.org/abs/1311.4883}{{\ttfamily
  arXiv:1311.4883 [hep-th]}}.

\bibitem{Bachas:2013vza}
C.~Bachas, E.~D'Hoker, J.~Estes, and D.~Krym, ``{M-theory Solutions Invariant
  under $D(2,1;\gamma) \oplus D(2,1;\gamma)$},''
  \href{http://dx.doi.org/10.1002/prop.201300039}{{\em Fortsch. Phys.}
  {\bfseries 62} (2014) 207--254},
  \href{http://arxiv.org/abs/1312.5477}{{\ttfamily arXiv:1312.5477 [hep-th]}}.

\bibitem{Bena:2023rzm}
I.~Bena, A.~Houppe, D.~Toulikas, and N.~P. Warner, ``{Maze Topiary in
  Supergravity},'' \href{http://arxiv.org/abs/2312.02286}{{\ttfamily
  arXiv:2312.02286 [hep-th]}}.

\bibitem{Sen:1995in}
A.~Sen, ``{Extremal black holes and elementary string states},''
  \href{http://dx.doi.org/10.1142/S0217732395002234}{{\em Mod. Phys. Lett.}
  {\bfseries A10} (1995) 2081--2094},
\href{http://arxiv.org/abs/hep-th/9504147}{{\ttfamily arXiv:hep-th/9504147}}.

\end{thebibliography}\endgroup

\end{adjustwidth}


\end{document}